\documentclass[12pt]{iopart}

\usepackage{iopams}  
\usepackage{amssymb}
\usepackage{graphicx}
\usepackage{setspace}
\usepackage{tocloft}
\usepackage{xcolor}
\usepackage{footmisc}
\usepackage{rotating,longtable}

\usepackage[]{hyperref}
\hypersetup{hyperindex=true}

\newcommand{\doclink}[2]{\href{#1}{#2}\footnote{\url{#1}}}

\begin{document}

\title[TANSPEC]{TANSPEC: TIFR-ARIES Near Infrared Spectrometer}

\author{Saurabh Sharma$^{a*}$, Devendra K. Ojha$^b$, Arpan Ghosh$^a$, Joe P. Ninan$^b$, Supriyo Ghosh$^b$, Swarna K. Ghosh$^b$, P. Manoj$^b$, Milind B. Naik$^b$,
Savio L. A. D'Costa$^b$, B. Krishna Reddy$^a$, Nandish Nanjappa$^a$, Rakesh Pandey$^a$, Tirthendu Sinha$^a$, Neelam Panwar$^a$, Susmitha Antony$^{b,c}$,
Harmeen Kaur$^d$, Sanjit Sahu$^a$, Tarun Bangia$^a$, Satheesha S. Poojary$^b$, Rajesh B. Jadhav$^b$, Shailesh B. Bhagat$^b$, Ganesh S. Meshram$^b$,
Harshit Shah$^b$, John T. Rayner$^e$, Douglas W. Toomey$^f$, and Pradeep R. Sandimani$^b$ }

\address{$^a$ Aryabhatta Research Institute of Observational Sciences (ARIES), Manora Peak, Nainital 263 001, India}
\address{$^b$ Tata Institute of Fundamental Research (TIFR), Homi Bhabha Road, Colaba, Mumbai - 400 005, India}
\address{$^c$ Indian Institute of Astrophysics (IIA), II Block, Koramangala, Bengaluru 560 034, India}
\address{$^d$ Center of Advanced Study, Department of Physics DSB Campus, Kumaun University Nainital, 263001, India}
\address{$^e$ Institute for Astronomy, University of Hawaii, 2680 Woodlawn Drive, Honolulu, HI 96822, USA}
\address{$^f$ Mauna Kea Infrared, LLC, 21 Pookela St.Hilo, HI 96720, USA}

\ead{saurabh@aries.res.in}
\vspace{10pt}
\begin{indented}
\item[]March 2022
\end{indented}

	\begin{abstract}
	We present the design and performance of the TANSPEC, a medium-resolution $0.55-2.5~\mu$m cryogenic spectrometer and imager, now in operation at the 3.6-m Devasthal Optical Telescope (DOT), Nainital, India.  The TANSPEC provides three modes of operation which include, photometry with broad- and narrow-band filters,  spectroscopy with short slits of 20$^{\prime \prime}$ length and different widths (from 0.5$^{\prime \prime}$ to 4.0$^{\prime \prime}$) in cross-dispersed mode at a resolving power R of $\sim$2750,  and spectroscopy with long slits of 60$^{\prime \prime}$ length and different widths (from 0.5$^{\prime \prime}$ to 4.0$^{\prime \prime}$) in prism mode at a resolving power R of $\sim$100-350.  TANSPEC's imager mode provides a field of view of 60$^{\prime \prime} \times 60^{\prime \prime}$ with a plate scale of 0.245$^{\prime \prime}$/pixel on the 3.6-m DOT.  The TANSPEC was successfully commissioned during April-May 2019 and the subsequent characterization and astronomical observations are presented here.  The TANSPEC has been made available to the worldwide astronomical community for science observations from October 2020.
	\end{abstract}

%
\vspace{2pc}

	\section{Introduction}
	\label{sect:intr}  

An instrument with simultaneous optical and near-infrared (NIR) spectroscopic/imaging capabilities on a 4-m class telescope will be an ideal observational tool needed for the efficient usage of the telescope time and can act as a work-horse for numerous science cases.
	Recently,  \doclink{https://aries.res.in/}{Aryabhatta Research Institute of Observational Sciences (ARIES)}, Nainital, India has installed a modern 3.6-m new technology optical 
	telescope (\doclink{https://aries.res.in/facilities/astronomical-telescopes/360cm-telescope}{3.6-m Devasthal Optical Telescope - DOT}), 
	the largest in India, at Devasthal (latitude = $29^\circ21^\prime39^{\prime \prime}.4$ N, 
	longitude= $79^\circ41^\prime03^{\prime \prime}.6$ E, altitude = 2450 m), Nainital in Uttarakhand.
	This facility is a major national observing facility in India
	and therefore it is of great advantage to complement this telescope 
	with an instrument having simultaneous optical-NIR coverage.
	
    The TIFR-ARIES Near Infrared Spectrometer (TANSPEC) is built in collaboration with 
	\doclink{https://www.tifr.res.in}{Tata Institute of Fundamental Research (TIFR)}, Mumbai, India;  
	ARIES, Nainital, India 
	and  \doclink{http://mkir.com/}{Mauna Kea Infrared LLC (MKIR)}, Hawaii, USA for the 3.6-m DOT.
	It is a unique spectrograph which provides simultaneous wavelength 
	coverage from $\sim$0.55 $\mu$m to 2.5 $\mu$m in spectroscopic  
	(cross-dispersed (XD) mode with $R\sim2750$ and prism mode with  $R\sim100-350$) as well as in imaging ($r^\prime i^\prime YJHK_s$ bands) modes.
	The importance of simultaneous optical-NIR observations
	has been realized in recent years for the Galactic and extragalactic astrophysical problems. The TANSPEC will be extremely sensitive to low temperature stellar photospheres (T$\sim$2500 K) 
	and objects surrounded by warm dust envelopes or embedded in the dust/molecular clouds. 
	It is therefore particularly suited to the study of low and very low mass stellar populations 
	(M/L dwarfs, brown dwarfs), strong mass-losing stars on the asymptotic 
	giant branch (AGB), young stellar objects (YSOs) still in their proto-stellar envelopes,  
	Gravitational lenses,  active galactic nuclei, etc.
	This instrument will also be ideal to produce a spectral library of a 
	large number of stars which can be used as training samples for the 
	future automated stellar classification tools.
	XD spectroscopy at resolving powers of $R\sim2750$ ($\sim$110 km/s) 
	with very broad spectral coverage (optical to NIR wavelengths) is very well suited to identify
	atomic and molecular features as well as to trace the high-speed dynamics such as stellar winds
	and broad-lined quasars. Deeply embedded YSOs and molecular species in protostellar disks, 
	comets, and AGB stars are best observed at $\sim 2-2.5$ $\mu$m. The low-resolution prism 
	mode is optimum for faint object spectroscopy such as brown dwarf spectral energy distribution, and solid-state features in asteroids and satellites. The long-slit
	in this mode is also useful to study extended objects such as planets and nebulae.
	
	At present, all the existing 2-m class telescopes within India (viz. \doclink{https://www.iiap.res.in/?q=iao.htm}{2.01-m Himalayan \textit{Chandra} Telescope} at Hanle, Ladakh; \doclink{http://www.iucaa.in/igo.html}{2-m  Inter-University Center for Astronomy and Astrophysics telescope} near Pune;  \doclink{https://www.iiap.res.in/?q=vbo.htm}{2.34-m Indian Institute of Astrophysics telescope} at Kavalur) have either optical or NIR imager/spectrograph instruments.
	The TANSPEC is only instrument in India and one of the few instruments  around the globe with simultaneous optical-NIR 
	capabilities.
	In this paper, we provide the details of the TANSPEC in Section~\ref{InstrumentOverview} (Overview), Section~\ref{OpticsSection} (Optics), Section~\ref{cryostat_section} (Cryostat), 
	Section~\ref{Electronics_and_control_system} (Electronics and Control system), and Section~\ref{SoftwareSection} (Software). The details on detectors and their performance are presented in Section~\ref{Detectors_and_its_performance}.
	The characterization and performances analyses of the TANSPEC instrument on the 3.6-m DOT are presented in Section~\ref{Performance_on_the_3.6-m DOT} and we conclude in Section~\ref{ConclusionSection}.

	\begin{figure*}
	\centering
	\includegraphics[width=0.85\textwidth, angle=0]{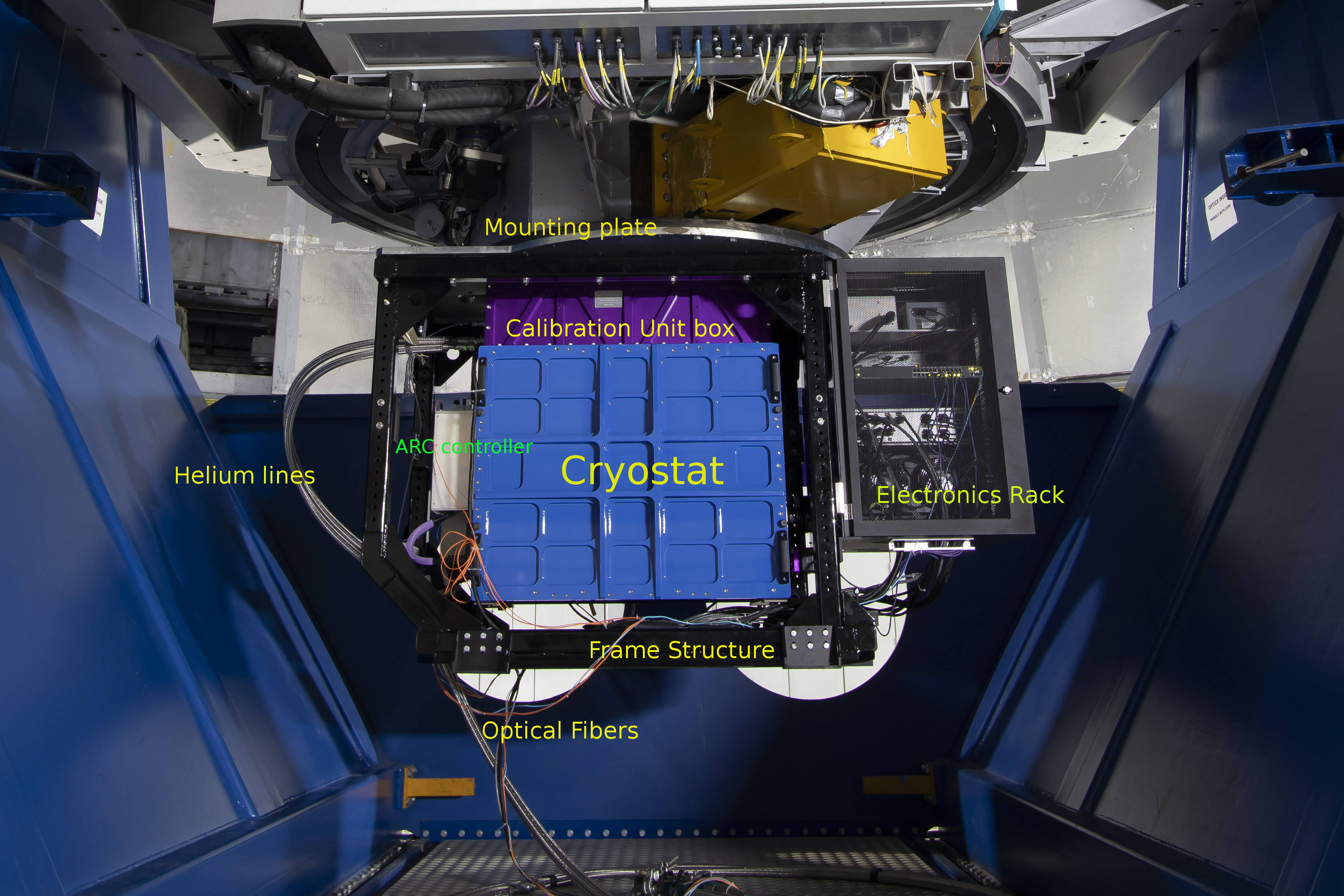}
	\caption{The TANSPEC mounted on the Cassegrain focus of the 3.6-m DOT. Various sub-components such as mounting plate, calibration unit box, cryostat, electronics rack, Helium lines,
	frame structure, and optical fibers of the TANSPEC instrument are also shown.
	The approximate size of this instrument (including electronic rack and mounting frame) is 2200 mm (length) x 1100 mm (width) x 1350 mm (height). The total weight of this instrument is 2000 kg, out of which, the instrument (blue+purple box) is 650 kg and mounting plate is of 350 kg. Remaining 1000 kg is added to the mounting frame to maintain the center of gravity to balance the telescope.}
	\label{tanspec}
	\end{figure*}

	\begin{figure*}
	\centering
	\includegraphics[width=0.85\textwidth, angle=0]{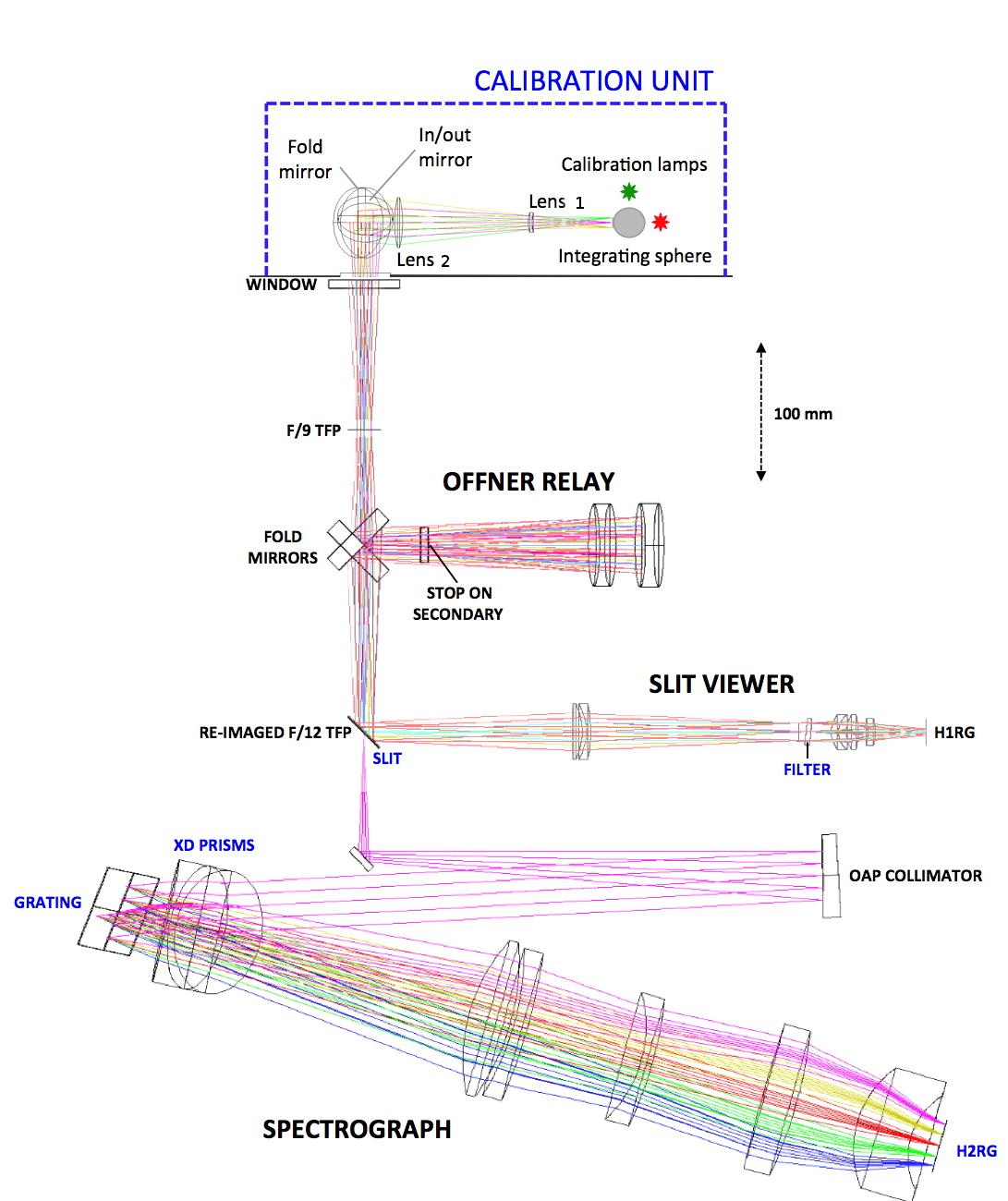}
	\caption {Overall optical layout of the TANSPEC comprising four major components: 
	Calibration unit, fore-optics (Offner relay), slit viewer and spectrograph. 
	The f/9 beam from the telescope comes to a focus at the telescope focal plane just inside the cryostat window. 
	A $60\times60$ arcsec$^2$ field of view is reimaged by an Offner relay onto the slit 
	where it is reflected into the slit viewer and re-imaged onto an  H1RG array 
	at 0.25 arcsec per pixel. A slot in the slit transmits the f/12 beam into the 
	spectrograph where it is dispersed and re-imaged onto an H2RG array. 
	In this view ray tracing for order three in the cross-dispersed mode is shown. 
	A warm calibration unit is located above the cryostat window.}
	\label{layout}
	\end{figure*}

	\section{Instrument overview} \label{InstrumentOverview}
	
  	A photograph of the TANSPEC mounted on the main port (Cassegrain focus) of the 3.6-m DOT through a rigid mounting plate is shown in Fig \ref{tanspec}. The cryostat (blue box) is interfaced with a rigid box (purple), which houses a spectral calibration unit consisting of transfer optics, integrating sphere, continuum and arc lamps and interfaces with the mounting plate of the instrument. The black frame in the photograph is the structure attached to the mounting plate which holds the counter-weights to balance this instrument and the electronics rack (black box). The electronics rack holds the power supplies, Ethernet power switches, temperature controller of the instrument. The photograph also shows the pair of Helium lines coming out from the instrument and going through the azimuth axis of the telescope
       to the ground floor of the telescope building and then to the closed-cycle cryo cooler's compressor.
       The optical fibers (orange cables) from the array controllers are going out to the  located at the telescope observing room.
      
       There are three sections which make up the
       cold optics inside the cryostat: the fore-optics, the infrared (IR) slit viewer/guider, and the spectrograph. 
	The initial concept of this instrument is drawn from the SpeX \cite{2003PASP..115..362R},  
	a medium-resolution 0.8-5.5 $\mu$m spectrograph and imager for the \doclink{http://irtfweb.ifa.hawaii.edu/}{NASA Infrared Telescope Facility (IRTF)} at Hawaii. 
       An overall view of the TANSPEC optical layout is shown in Fig \ref{layout}.
       Briefly, the f/9 beam from the telescope or calibration lamp comes to a focus at the telescope focal plane (TFP)
       about 100 mm inside the CaF2 window.
       The TFP is then reimaged at f/12 onto slits in the slit wheel by an Offner relay
       containing two concave mirrors and one convex secondary mirror.
       The slit mirrors are tilted at 45 degrees to the incoming f/12 beam and
       reflect it to the slit viewer for imaging and guiding using   $1024\times1024$ pixels HAWAII-1RG (H1RG) array.
       The f/12 beam entering the spectrograph through the slot in the slit is first folded and then collimated on to a dispersing optics.
       The dispersed beam is focused onto a $2048\times2048$ pixels HAWAII-2RG (H2RG) array by a set of camera lenses.
       All the lenses are anti-reflection coated for 0.55-2.5 $\mu$m wavelength range.
       This was a challenge because the wavelength range is broad and encompasses the visible and NIR range. 
       We were able to achieve designs that gave around 2\% reflectivity or less over the full wavelength range. 
       As a typical visible wavelength filters are designed for CCDs with a $\sim$1 $\mu$m wavelength cutoff,
        the filter used in the slit-viewer are specially designed to block out of band wavelengths upto 2.7 $\mu$m. 
	This is required also because there is much higher background flux levels at the 1-2.5 $\mu$m range.
       Individual Astronomical Research Cameras (ARC) array controllers run each of the spectrograph and imaging arrays and are mounted on the cryostat.
       The array controller  uses a single-board SPARC computer and  four digital signal processor
       (DSP) boards.
       A graphical user interface (GUI) of the TANSPEC runs on a Linux workstation placed in the observing room connected to
       the controllers of the arrays  by the optical fibers.

	\section{Optics} \label{OpticsSection}	
	
	\subsection{Calibration Unit}

	A calibration unit is used for the flat fielding and the wavelength calibration of the raw spectra taken through this spectrograph.
	As shown in Fig \ref{layout}, it has an assembly of calibration lamps, integrating sphere, camera lenses, and
        a fold mirror (mounted on a linear translation stage that moves into the telescope beam and picks off the
        beam from the calibration lamps when required). 
	The pick-off mirror and lamp control can be remotely operated from the instrument software.
	The calibration optics are designed in a way
        to match the telescope's f/number and to evenly illuminate the TFP inside the TANSPEC with a 15 mm
	diameter flat field. This is the size required to cover the longest slits (60 arcsec). A 12.5 mm 
	diameter exit port (field stop) of an integrating sphere serves as the uniform (Lambertian) flat field. This field is magnified
	to 15 mm diameter by a 246 mm focal length lens (Lens 1) placed 82 mm from the exit port. A 125 mm focal
	length CaF2 lens (Lens 2) re-images this field at 1:1 onto the f/9 TFP inside the cryostat. At
	the same time, Lens 2 images the 15 mm aperture stop of the Lens 1 onto the telescope secondary mirror to
	create an exit pupil identical to that of the telescope so that the calibration light sources illuminate the
	instrument identically to the telescope. The root mean square (RMS) spot diameter in the TFP from the calibration optics is
	about 0.5 mm. This is sufficient enough to satisfy the uniformity requirement of better than 1\% across the
	slits, which is limited by the integrating sphere. The calibration unit has \doclink{https://www.newport.com/f/pencil-style-calibration-lamps}{Argon and Neon Discharge lamps} and two
	Tungsten continuum (of different brightness) lamps which cover the entire wavelength range (0.55-2.5 $\mu$m) of the spectrograph.

	\subsection{Fore Optics}

\begin{figure*}
	\centering
	\includegraphics[width=0.55\textwidth, angle=0]{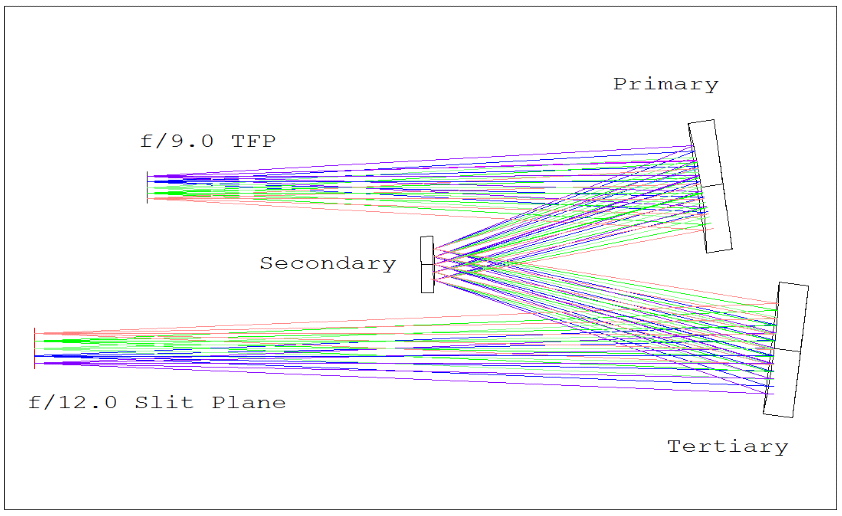}
	\caption{The fore-optics re-image the TFP onto the slit plane. This is done with an Offner relay. In this view the fold mirrors are
		removed and the slit plane is shown un-tilted. The telescope pupil of 14.3 mm diameter 
	is imaged onto the Offner secondary mirror. A cold stop is placed here.  }
	\label{offner}
	\end{figure*}

	The function of the fore-optics is to re-image the TFP onto the slits with the necessary magnification required
	for the spectrograph. At the same time an image of the entrance pupil needs to be formed where a cold stop
	can be located to baffle thermal flux and off-axis scattered light from the telescope. An optimized cold stop
	cannot be located in the spectrograph since the slit diffractively blurs the entrance pupil image. In the TANSPEC,
	this is achieved with an Offner relay (cf. Fig \ref{offner}), which consists of three on-axis spherical mirrors. 
	The primary  and secondary of the offner relay consists of two concave and one convex mirrors, respectively.  
	The two mirrors of the primary are a little different in focal length so that we can convert from f/9 beam to f/12 beam 
	which is the desired F-number of the beam at the slit entrance to the spectrograph optics.  
	This results in a slight reduction of the Offner optical performance that is not significant for the system performance. 
	The Offner-relay is shown in Fig \ref{offner} with the folding flat mirrors removed. 
	It re-images the f/9 $60^{\prime \prime}\times60^{\prime \prime}$ TFP onto the slit plane at f/12. 
	We painted the center of the cold secondary of Offner relay (onto which telescope pupil is imaged) black to work as a cold stop. Being all reflecting, the
	Offner relay is completely achromatic and has high and constant throughput across the desired wavelength
	range.

	\subsection{Slit viewer}

\begin{figure*}
	\centering
	\includegraphics[width=0.49\textwidth, angle=0]{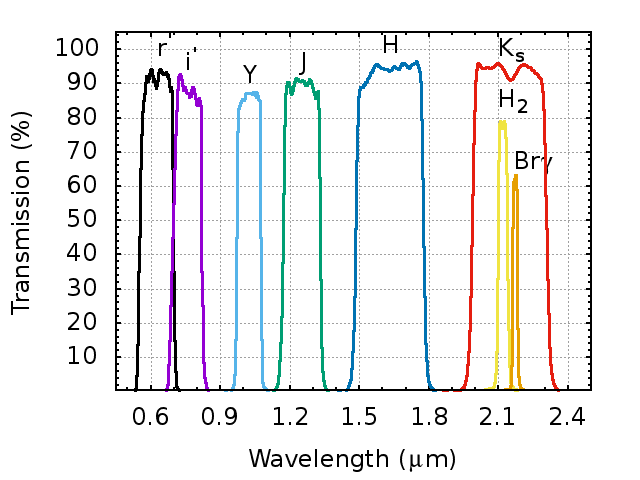}
	\caption{Response curves for the filters used in the slitviewer of TANSPEC.}
	\label{ffil}
	\end{figure*}

	The IR slit viewer allows  spectra of very red and embedded objects through NIR guiding.
	It is also a fairly capable imager. 
	The slit viewer is a collimator-camera re-imaging system consisting of BaF2-LiF collimator and LiF-BaF2-ZnSe
	camera lenses that re-image the  $60^{\prime \prime}\times60^{\prime \prime}$ 
	slit plane onto H1RG array
	with an image scale of $\sim0.25^{\prime \prime}$ (cf. Fig \ref{layout}). 
	For Nyquist sampling of the field of view (FOV) reflected of the slit, only the central $480\times416$ pixel$^2$ area (starting from x=416 and y=288 pixels) is used.
	The collimator doublet forms another image of the system entrance pupil where slit-viewing 
	filters are placed (each 25.4 mm diameter). This second stop is not optimized to cold baffle since
	the cold stop in the fore-optics does this. However, it is the best location for the imaging filters (small and
	uniform pupil image).
	The slit viewer has a filter-wheel with 12 slots to accommodate  filters of different wavelengths. This wheel have position accuracy of 50 $\mu$m.
	The details of the filters-wheel and filters  used in the TANSPEC slit-viewer are given in Table \ref{wheel} and Table \ref{fil}, respectively. 
	The $Y,J,H,K_s$ filters are in Mauna Kea Observatory (MKO) system \cite{2002PASP..114..180T}
	make by \doclink{https://www.asahi-spectra.com/}{Asahi Spectra, Japan}. 
	The $r^\prime ,i^\prime ,H_2,$ and $Br\gamma$  filters are made by \doclink{https://www.andovercorp.com/}{Andover Corporation, USA}, where the $r^\prime ,i^\prime $ 
	filters are in the Sloan digital sky survey (SDSS) system \cite{1979PASP...91..589B}.
	The response curves for all the filters are provided in Fig \ref{ffil}.

        \begin{table*}
        \centering
        \caption{\label{wheel}The details of the  filter and slit wheels.}
        \begin{tabular}{@{}c|cc|ccc@{}}
        \hline
        Position &  \multicolumn{2}{c}{Filter wheel}       &  \multicolumn{3}{c}{Slit wheel} \\
                  & Filter     & Note & Slit width ($^{\prime\prime}$) & Slit length ($^{\prime\prime}$) & Note  \\
        \hline
        1     &$J$      &Broad-band filter    & 0.5   &       20      &XD mode       \\
        2     &$H$      &Broad-band filter    & 0.75  &       20      &XD mode       \\
        3     &$K_s$      &Broad-band filter    & 1.0   &       20      &XD mode       \\
        4     &$Y$      &Broad-band filter    & 1.5   &       20      &XD mode       \\
        5     &$i^\prime $     &Broad-band filter    & 2.0   &       20      &XD mode       \\
        6     &$r^\prime $     &Broad-band filter    & 4.0   &       20      &XD mode       \\
        7     &$Br\gamma$&Narrow-band filter    & 0.5   &       60      &Prism mode  \\  
        8     &$H_2$  &Narrow-band filter    & 1.0   &       60      &Prism mode    \\
        9     &Blank1 & -    & 2.0   &       60      &Prism mode    \\
        10    &Blank2 & -    & 4.0   &       60      &Prism mode    \\
        11    & Free slot        & -    & Mirror&        -      &Imaging mode  \\
        12    &Pupil  & -    & Pupil &        -      & -            \\
        \hline
        \hline
        \end{tabular}
        \end{table*}

        \begin{table*}
        \centering
        \caption{\label{fil}Specification of the filters used in the slit-viewer of TANSPEC.}
        \begin{tabular}{@{}cccc@{}}
        \hline
		Filter  Name	&	Central Wavelength	& Wavelength range$^a$ 	& Photometric	\\
				&	($\mu$m)		&	($\mu$m)	& system	\\	
        \hline
		$r^\prime $		&0.612				&0.555-0.670			&SDSS \\	
		$i^\prime $		&0.744				&0.681-0.805			&SDSS \\	
		$Y$		&1.020 				&0.970-1.070		&MKO 	\\	
		$J$		&1.250 				&1.170-1.330		&MKO   	\\	
		$H$		&1.635 				&1.490-1.780		&MKO  	\\	
		$K_s$		&2.150 				&1.990-2.310		&MKO   	\\	
		$H_2$		&2.122 				&2.097-2.138		&-     	\\	
		$Br\gamma$	&2.171 				&2.160-2.181		&-     	\\	
        \hline
        \end{tabular}\\
	$^a$: at 50\% transmission of the peak\\
        \end{table*}

	\subsection{Spectrograph Unit}

	\subsubsection{Slits}

\begin{figure*}
	\centering
	\includegraphics[width=0.45\textwidth, angle=0]{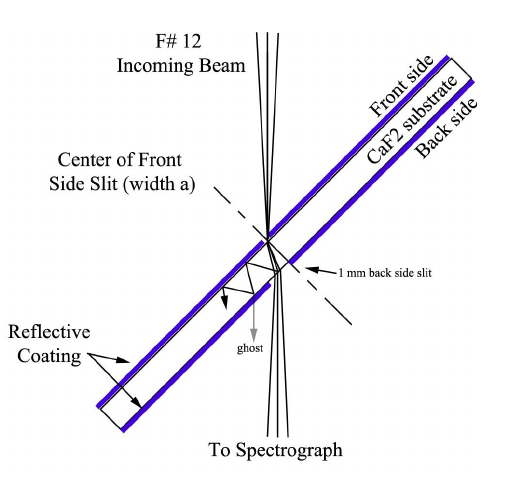}
	\caption{Description of a slit used in the TANSPEC instrument.}
	\label{fslit}
	\end{figure*}

    The spectrograph unit has a slit-wheel with 12 slots to accommodate  slits of different sizes and have a position accuracy of 50 $\mu$m.
	The details of the slits are listed in Table \ref{wheel} and is shown in  Fig \ref{fslit}.
	The slits are 1.00 mm thick Fused Silica substrates that
	have a silver-coated mirror with a slot lithographically
	applied to the front surface. The slit mirror is oriented
	at 45 degrees to the incoming f/12 beam and reflects a
	60 $\times$ 60 arcsec$^2$ FOV into the slit viewer. Since the slit is
	applied lithographically, it is very sharp and any
	scatter from the slit edges is minimized. To avoid the
	ghost reflection at the back surface of the slit
	substrate entering the spectrograph, an absorbing
	surface is applied to the back surface of the slit
	substrate. This back surface has a suitably positioned
	slot to transmit the primary beam and absorb the ghost. 
	The slit mirrors were coated with Aluminium and were made by
    \doclink{https://www.photo-sciences.com}{Photosciences Inc, USA}.

	\subsubsection{XD mode}
	
\begin{figure*}
	\centering
	\includegraphics[width=0.65\textwidth, angle=0]{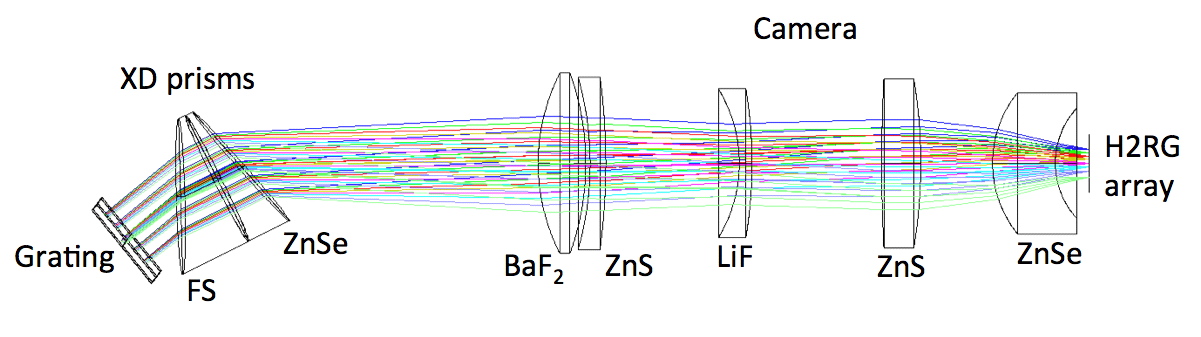}
	\includegraphics[width=0.33\textwidth, angle=0]{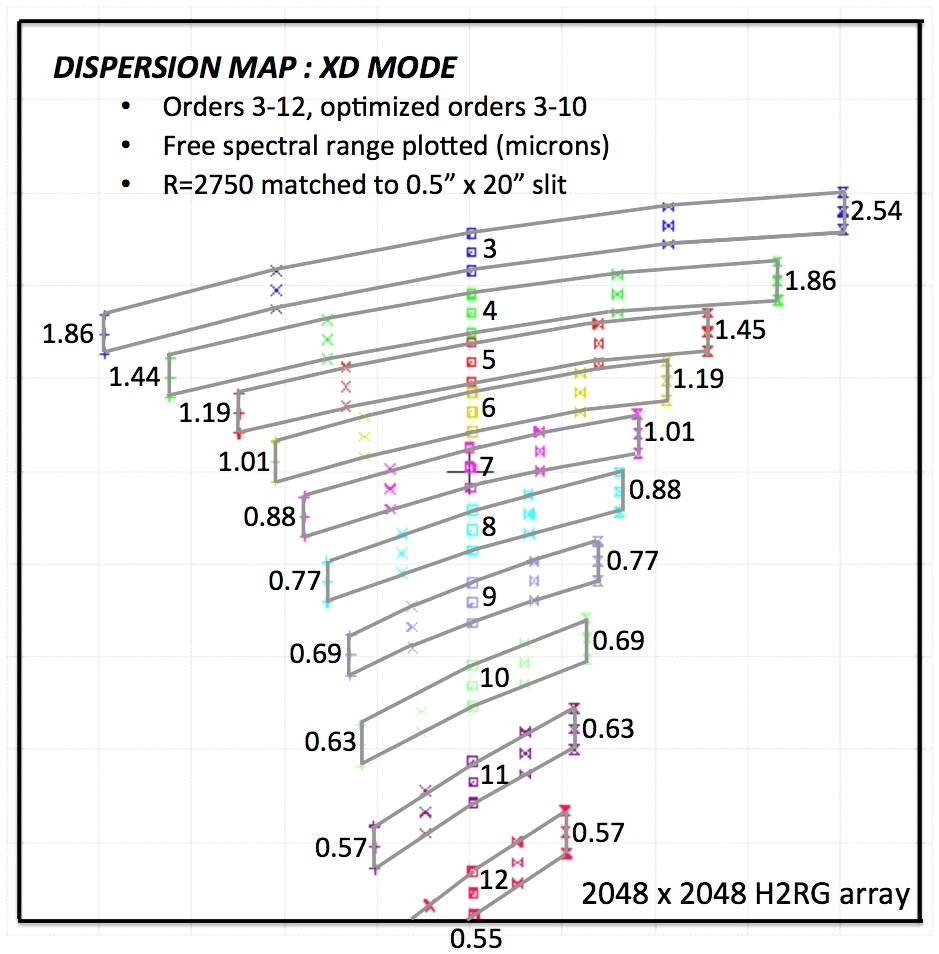}
	\caption{Left panel: Ray diagram of the spectrograph XD mode for orders 3-10.
	 Right panel: Dispersion map in the XD mode optimized for 2.5-0.63 $\mu$m (orders 3-10). Orders 11-12 are covered but optical aberrations limit R$<$2500. 
	The H2RG array is sensitive between $\sim$2.5-0.4 $\mu$m.}
	\label{fray}
	\end{figure*}

	The f/12 beam entering the spectrograph at the slit is first folded and then collimated by an Aluminium off-axis-parabolic 
	(OAP) mirror. The OAP and its coatings were done by \doclink{https://dpo.webspace.durham.ac.uk/}{Durham University, Precision optics lab, UK}. 
	 An image of the entrance pupil (35 mm diameter) is formed at a distance of one focal length
	from the OAP (420 mm). Grating and prism dispersing elements are located here, housed in the grating wheel. 
	One position of the wheel houses the grating and cross-dispersing double prisms (the XD mode),
	and the other houses a mirror and single prism (the low-resolution mode). The dispersed beam is focused
	onto  H2RG array by a five-element refractive camera lens: BaF2-ZnS-LiF-ZnSe-ZnSe. The optical layout and
	spectral format  of the XD mode are shown in Fig \ref{fray}.  This mode is optimized for wavelengths
	0.63-2.54 $\mu$m (orders 3-10). To minimize aberrations, these orders are centered on the array and the
	geometric (untoleranced) resolving power is R$\sim$2750 at the center of each order. The H2RG array is sensitive
	from 0.4  $\mu$m to 2.5  $\mu$m approximately, therefore wavelengths 0.63 - 0.55  $\mu$m (orders 11-10) are
	also imaged onto the array but at slightly reduced resolution due to higher optical aberrations and minor vignetting in	the camera lenses.
	
	The configuration for the XD mode of the TANSPEC is given in Table \ref{grating}.
	The grating used in the TANSPEC is an off-the-shelf master grating (Aluminium substrate) made by \doclink{https://www.newport.com/b/richardson-gratings}{Richardson Gratings, USA,} closely approximating the derived design. 
	Fused Silica and ZnSe prisms are used in double-pass to achieve cross dispersion 
	over the desired spectral range of $\sim$0.55-2.5 $\mu$m.
	The prisms are vacuum spaced. There is a slight angle introduced into the second prism so that the two inner surfaces are at an angle to each other.  This prevents fringing caused by the two surfaces being parallel.
	One disadvantage of prisms is that they usually have lower dispersion than gratings. 
	Using the prisms in double pass configuration helps to compensate for this. 
	Also, by using prisms made from Fused Silica and ZnSe the order separation 
	can be made almost constant in wavelength over the range $\sim$0.55-2.5 $\mu$m 
	leading to a much better use of the detector format than is possible for a single prism or 
	grating (order separation is proportional to $\lambda^2$).

        \begin{table*}
        \centering
        \caption{\label{grating}TANSPEC near-Litrow configuration in the XD mode.}
        \begin{tabular}{@{}cc@{}}
        \hline
	Parameter	&	Value\\
	\hline
	R				&2750\\
	Slit width 			&0.5 arcsec (2 pixel sampling)	\\
	Slit length			& 20 arcsec in XD mode\\
	Collimated beam diameter	& 35.0 mm\\
	Grating blaze angle		& 17.5 degrees\\
	Grating groove frequency	&90 lines/mm\\
	Grating blaze wavelength (m=1)	& 6.7 micron\\
	Angle between incident and diffracted beams&9.5 degrees\\
	ZnSe prism apex angle		& 15 degrees\\
	FS prism apex angle		& 25 degrees\\
	Input beam			& f/12\\
	OAP focal length		&420 mm\\
	OAP off-axis angle		& 5.5 degrees\\
	Camera focal length		&144 mm\\
        \hline
        \hline
        \end{tabular}
        \end{table*}

	\subsubsection{Prism mode}

\begin{figure*}
	\centering
	\includegraphics[width=0.68\textwidth, angle=0]{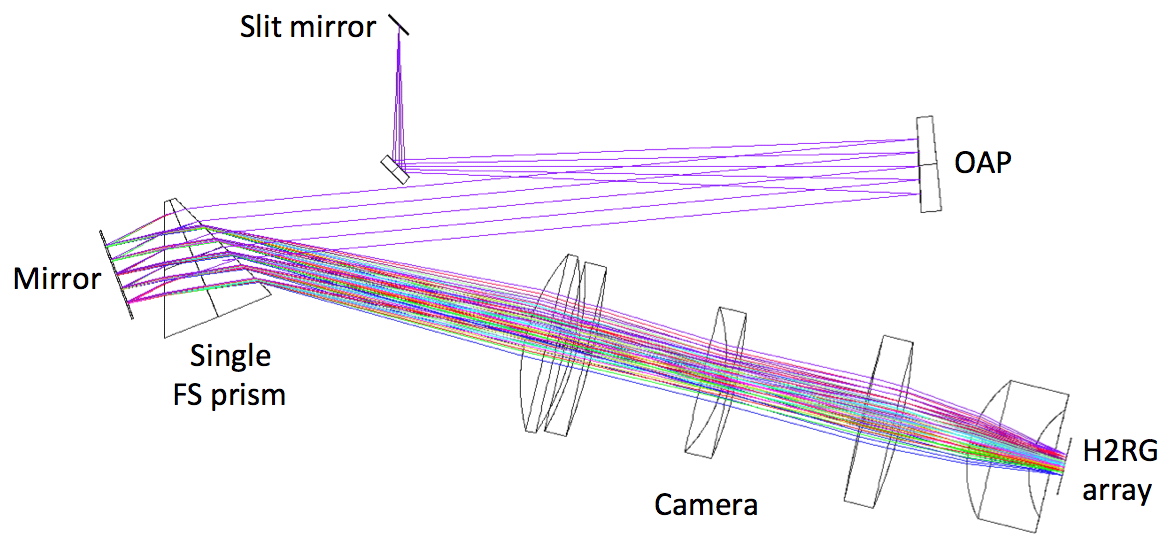}
	\includegraphics[width=0.30\textwidth, angle=0]{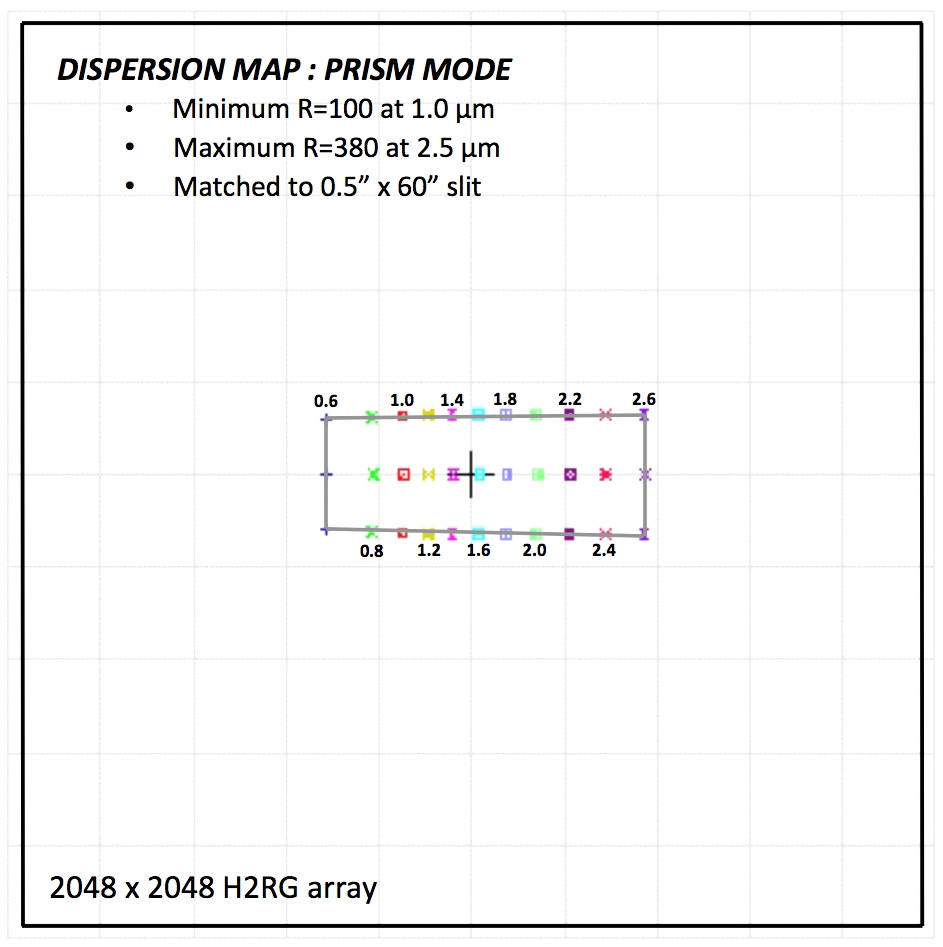}
	\caption{Left panel: Ray diagram for the prism (low-resolution) mode. Right panel: Dispersion map covering 0.55 - 2.5 $\mu$m wavelength range in the prism mode. }
	\label{fprismm}
	\end{figure*}

	In the prism mode, the grating and double prism are replaced by a mirror and single prism  (cf. Fig \ref{fprismm}).
	The prism is mode of fused silica and is of size similar to the cross-dispersing prisms.
	The prisms and coatings on the prisms were done by \doclink{https://www.rr-optics.com}{Rainbow Research Optics INC}. 
	The dispersing elements are mounted in a grating wheel. All the other optics are unchanged. 
	Consequently, stray light levels in the prism mode are very similar to those in the XD mode.
	The apex of the single Fused Silica prism is oriented at 90 degrees to apex of the double prisms of the XD mode so
	that dispersion is along array rows (in the XD mode, dispersion is along array columns). The resulting dispersion map is 
	also shown in Fig \ref{fprismm}. Due to the change in
	refractive index of Fused Silica with wavelength, R changes by a factor of three across the wavelength range.
	The prism is
	designed for R$\sim$100 matched to the narrowest slit (0.5$^{\prime\prime}$)  at about 1.0 $\mu$m and is higher at shorter and longer wavelengths. 
	A complication of the design is that the angle
	required to prevent the input and exit beams to the prism from colliding, creates anamorphic shrinking
	in the dispersion direction. This results in the low-resolution mode 0.5$^{\prime\prime}$ slit being sampled by around three
	pixels instead of two pixels as in the XD mode (where the anamorphic effects are in the XD direction).
	
	\subsubsection{Stray Light and Baffling}

	A cold Lyot stop in the fore-optics is used to control the thermal background and  stray light  from the telescope and sky. The	light paths in the instrument 
	are enclosed mostly in baffled tubes in  Aluminium enclosures. The enclosures are designed in a way to prevent 
	thermal background and stray light from the cryostat components which are warmer than $\sim$80 K (e.g., drive shafts, wires, warm vacuum jacket etc.) 
	that might scatter along the path of the light towards the detectors. The aim of the design was to keep  instrument background below the dark
	current of the detector ($\sim$0.1 electron/sec).  Because of the broad wavelength coverage of the TANSPEC, large
	variation in either sky or object brightness can occur across the spectra. 
	Therefore, any ghost reflection from the brighter part of the spectrum to the fainter parts can 
	degrade the quality of the data. The main areas of the concern are narcissus ghosts from the prisms and camera lens. 
	The ray-tracing of the former  suggests that these ghosts can be minimized by tilting the
        H2RG array by 1 degree (this is enough to redirect any ghost reflections off the prisms and incurring no
        measurable defocus across the array). The latter are extremely out of focus and it does not pose any concern. 
	The amount of ghosting in the TANSPEC is of the order of few percent and can be removed by sky subtraction.

	\begin{figure*}
	\centering
	\includegraphics[width=0.46\textwidth, angle=0]{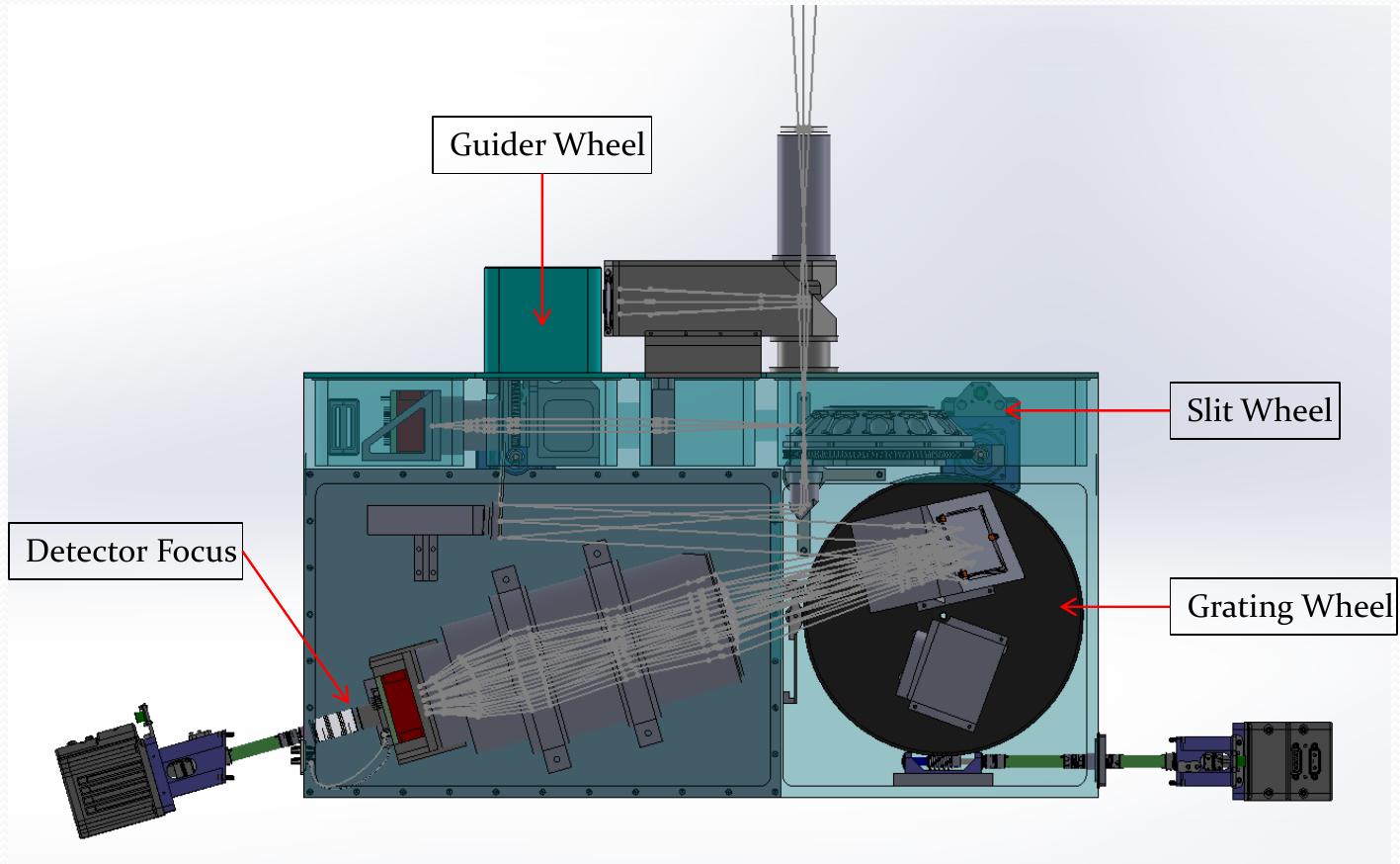}
	\includegraphics[width=0.44\textwidth, angle=0]{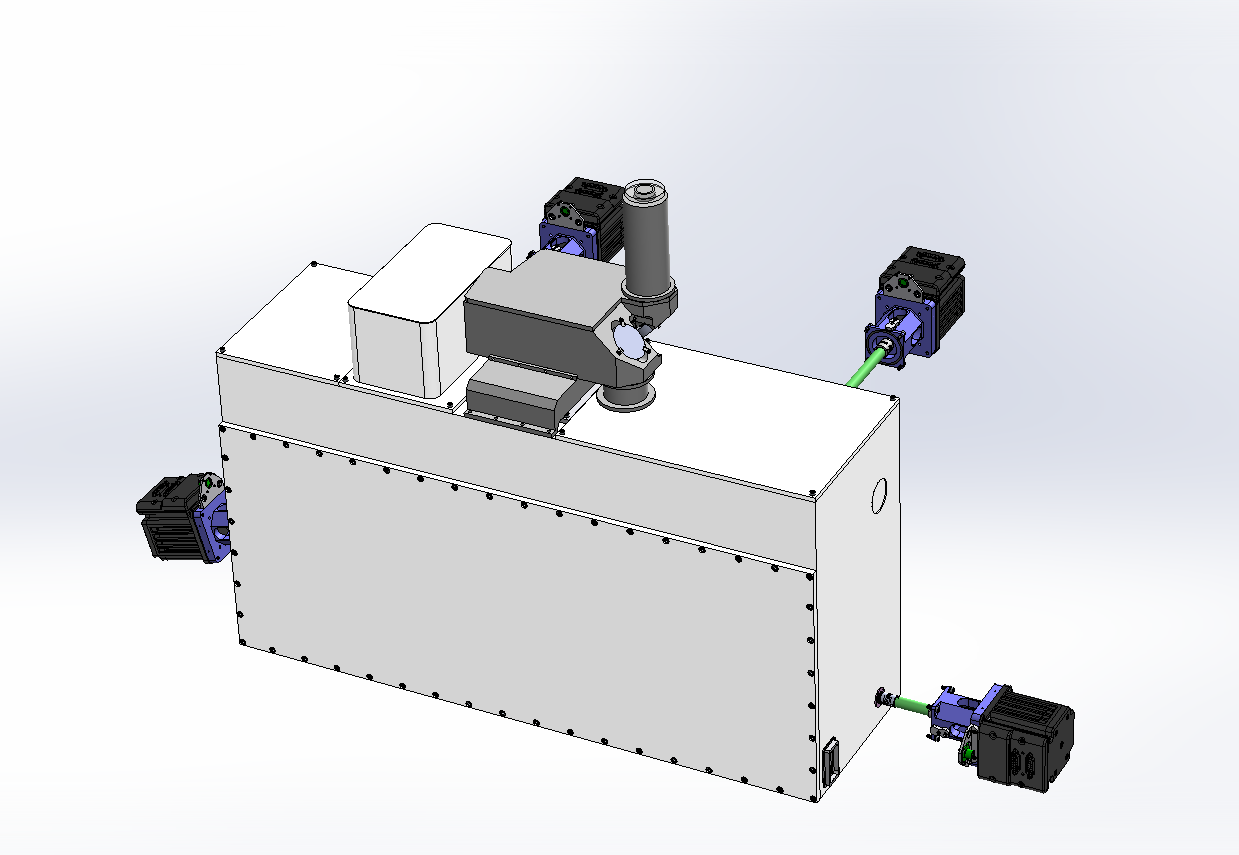}
	\includegraphics[width=0.45\textwidth, angle=0]{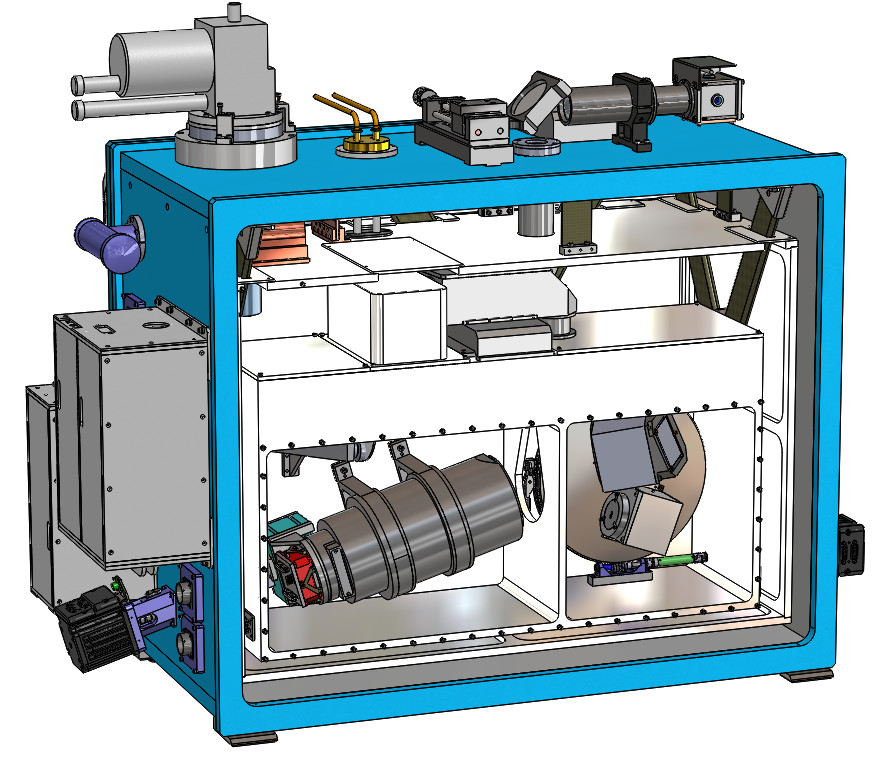}
	\includegraphics[width=0.45\textwidth, angle=0]{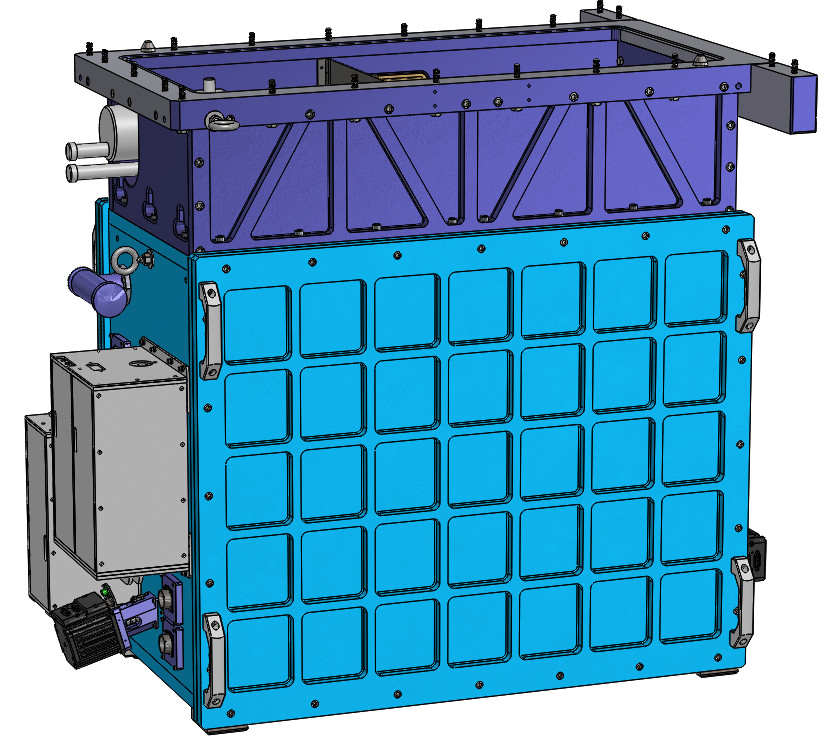}
	\caption{Mechanical assembly of the TANSPEC. The various mechanisms along with ray trace (top left), cold structure (top right), 
	vacuum jacket (blue box), radiation shield, cold structure along with other components (bottom left), and the entire cryostat assembly (bottom right)
	along with top box (purple box containing calibration unit, LN$_2$2 fill port, cold head) are shown.}
	\label{fmech}
	\end{figure*}

	\section{Cryostat} \label{cryostat_section}

	\subsection{Mechanical structure}

	The entire mechanical assembly of the TANSPEC is shown in Fig \ref{fmech}.
	The TANSPEC instrument has an envelope size of 1900 mm (width) x 915 mm (length) x 1148 mm (height)
	which is within the telescope instrument envelop.
	The  weight of the TANSPEC and its mounting plate are 650 kg and 350 kg, respectively.
	Therefore, to balance this instrument and maintain the center-of-gravity of the 3.6-m DOT, 
	an extra frame with counterweights (1000 kg) on it is added to the instrument. 
	Electronics cabinet containing power-supplies, 

	\setcounter{footnote}{0}
	\doclink{https://www.lakeshore.com/products/categories/overview/temperature-products/cryogenic-temperature-controllers/model-336-cryogenic-temperature-controller}{Lakeshore, USA temperature controller} and \doclink{https://www.wti.com/collections/power-transfer-switch}{Western Telematic Inc. (WTI), USA Ethernet power switches} are also mounted on this frame. The TANSPEC is attached to the telescope 
	through a mounting plate via 
	a rigid interface box. This box is mounted in between the top plate and the vacuum jacket.  A calibration unit is placed inside this
	box. The  box is 280 mm deep and the walls are of 25 mm thickness. The top plate of the vacuum jacket
	contains the liquid nitrogen fill port, entrance window,  and the mount for closed-cycle cooler. 
	The TANSPEC along with its frame mounted on the 3.6-m DOT is shown in Fig \ref{tanspec}.  

	The vacuum jacket of the TANSPEC has a four-sided center section
	weighing about 205 kg and is bolted by two large O-ring sealed
	covers, each weighing about 38 kg. 
	The four-sided center section is made from four plates of
        6061-T6-Aluminium which are  electron-beam welded together. The vacuum jacket  has a external dimensions of
	1070 mm (length) x 860 mm (height) x 630 mm (width). 
	The vacuum jacket has no hard connections except at the
	rigid top plate (25 mm thick).
	Therefore, if we can keep the flexures below the yield strength of the Aluminium
        along with the flexure of shafts (connected to the warm
        motors mounted outside of the vacuum jacket) within the specifications,
	we can achieve a very stable configuration by using a thick vacuum
	jacket walls of 25.4 mm. Although, the large vacuum jacket covers are with ribs that taper up to 70 mm. 
 	All the internal surfaces are 
	mirror finished to provide good radiation shielding and to improve the liquid Nitrogen (LN$_2$) hold time. 
	Inside this there is an radiation shield which is closed-cycle cooler cooled shield that removes most
	of the thermal radiation from the vacuum jacket. The wall of the LN$_2$ cooled cold bench/box is 12.5 mm thick
	and hangs inside the vacuum jacket through fiber-glass V-trusses (3) which are connected to the top plate inside the radiation shield. 
	The detectors are thermally clamped to the cold structure.
	The radiation shield is mounted around cold structure through four shear webs  connected to the top of the vacuum jacket.

	\subsection{Mechanisms}

	The TANSPEC has mechanisms for the slit wheel, guide channel filter wheel, grating wheel, spectrograph
	channel detector focus and calibration unit pick-off mirror. 
	It uses external warm smart motors manufactured by \doclink{https://www.animatics.com/}{Animatics Corporation, USA.} mounted on one of the two
        covers of the vacuum jacket and are coupled to the wheels
        through ferrofluidic vacuum feed-throughs. The optics and detectors can be accessed by unbolting
	the opposite covers. Thees is no need to disassembled  motors for majority of the
	troubleshooting inside the cryostat.  The array controllers are mounted on the outer surface of  the cryostat. 
	The mechanisms have three different categories, i.e., continuously variable wheels, discrete
	position wheels with detents,  and a linear flex stage. 
	For the position sensing, we have used the non-contact Hall effect devices 
	similar to the SpeX instrument \cite{1998SPIE.3354..468R}. 
	All the wheels are equipped with 
	a stationary Hall effect sensor.
	These sensors are used to  encod wheels position  by
	placing the magnets at their edges. Opposite polarities
	signifies the homing and other positions. 
	The Hall effect sensors outputs  are passed through the comparator circuits and function like limit switches.
	Stepper motors of 400 steps/revolution in the worm-gear mechanisms is used to drive all the wheels with detented positions
	and the final position is determined by counting the steps taken from the home position.
	The H2RG array focus stage consists of an Aluminium alloy flex stage with anti-backlash linear position mechanism which provides $\pm$2.5 mm of focus travel
	through a screw coupled with a stepper motor. 
	The calibration lamp assembly has a plane mirror which can be moved 
	by using an motor and linear stage arrangement.

	\subsection{Cryogenic design and performance}

	The required cooling of this instrument is achieved through a hybrid system with closed-cycle cooling for the radiation shields and LN$_2$ cooling for the optical bench and detectors. 
	In order to keep the instrument background below the HxRG detector dark current ($\sim$0.1 electron/sec), LN$_2$ cooling of the cold structure is done.
	The radiation shield which surrounds the cold structure is cooled by using a closed-cycle cooler.
	A cooling temperature of $\sim$150 K is sufficient for the radiation shield to reduces the
        radiation load on the cold structure which can be managed very easily by the LN$_2$ can.
	The surface areas of the  cold structure, radiation shield and the vacuum jacket are 1.8 m$^2$, 2.6 m$^2$, and 4.3 m$^2$, respectively. 

	We have uses a \doclink{https://www.idealvac.com/files/manuals/Brooks9600High_1.pdf}{CTI model 9600 high voltage compressor} manufactured by \doclink{http://www.brooks.com}{Brooks, USA} along with two long (40 meters each)  pressurized Helium lines to cool the
	highly polished 5 mm thick 6061-aluminum radiation shield to less than 150 K.
	The estimated heat load on the cold structure consists of 2.1 W heat from the radiation shield having temperature of 150 K, 
	1.9 W from the holes in it (entrance window etc), 1.7 W from support V-trusses,
	1 W from the wiring, and 0.25 W from the motor shafts.
	The expected variation in the performance of the closed-cycle cooler does not
         effect the temperature of the cold structure as its heat load is negligible.
	The temperature variation due to the boil-off of the LN$_2$ ($\sim$0.5 K) is also
	small to affect the instrument performance (e.g., instrument background or focus). 

	The instrument can cool-down in $\sim$60 hours and the hold time for the 14.16 L capacity LN$_2$ can is $\sim$100 hours within the ambient temperature range of 283 to 301 K.
	Optical bench and guide detector are maintained at $76\pm 2$ K. 
	Warm-up also takes 60 hours, but by bowing out the LN$_2$ by circulating dry air,
	the cryostat can be warmed up faster .

	\begin{figure*}[h]
	\centering
	\includegraphics[width=0.49\textwidth, angle=0]{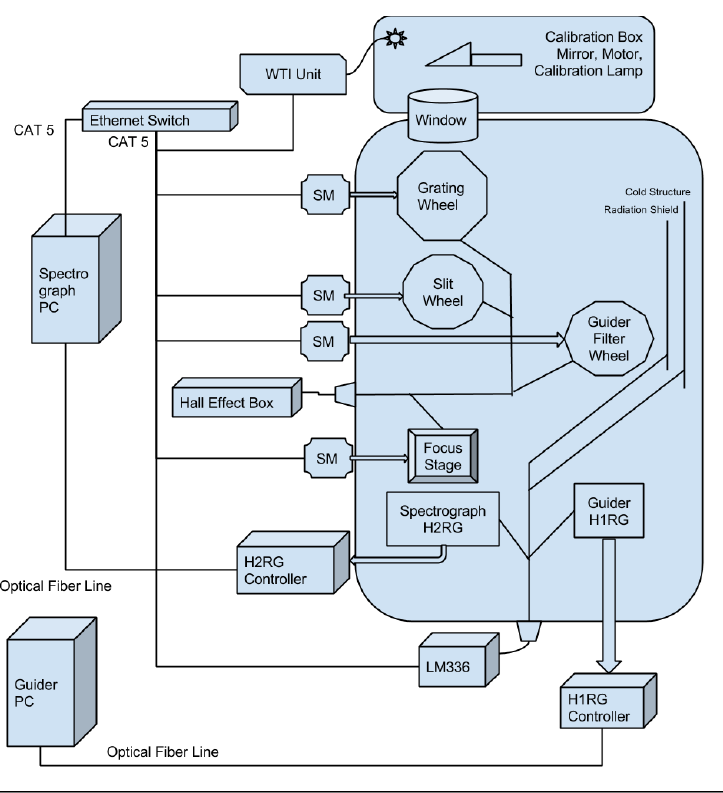}
	\caption{The representative scheme of the TANSPEC Hardware. Apart from a PC (for both guider and spectrograph), all other components are mounted on the telescope either on electronics rack or on the cryostat of the TANSPEC. }
	\label{fhardware}
	\end{figure*}

	\section{Electronics and Control System} \label{Electronics_and_control_system}

	Overall scheme of the TANSPEC hardware is shown in Fig \ref{fhardware}.
	All instrument electronics along with the instrument are mounted at the telescope Cassegrain port. 
	It consists of the array controllers for
	the two detectors, preamps, analog-to-digital (A/D) boards, DSP boards, remote Ethernet power control of
	all elements, mechanism drivers, temperature control monitoring of all parameters, Ethernet switch, power
	supplies, etc., which are controlled by the Unix Workstations placed in the control room and connected
	through optical fibers. The wheels, H2RG focus mechanics and calibration box mirror are moved using independent smart
	motors. The motors are controlled through a
	standard Ethernet interface. 
	Electronics cabinet is insulated and designed for glycol cooling to ambient temperature using glycol chillers.
	The detector arrays are controlled by  ARC Gen III controllers (H2RG: 4 channels and H1RG: 1 channel) from \doclink{http://www.astro-cam.com/Gen3Products.php}{ARC, Inc., USA}. It is also known as Leach Controller or  San Diego State University (SDSU) Controller. 
	The configuration of the arrays and its controllers are similar to our previous instrument `TIRSPEC' (see for details, \cite{2014JAI.....350006N}).
	All the  temperatures (arrays, radiation shield and cryostat) are monitored using a Lakeshore temperature controller (Model No: 336). 
	The electrical connections and the chassis
	for the spectrograph controller have provisions to increase the channels to 32 in case 32 output mode of the
	H2RG is required at a later stage.

	\begin{figure*}
	\centering
	\includegraphics[width=0.65\textwidth, angle=0]{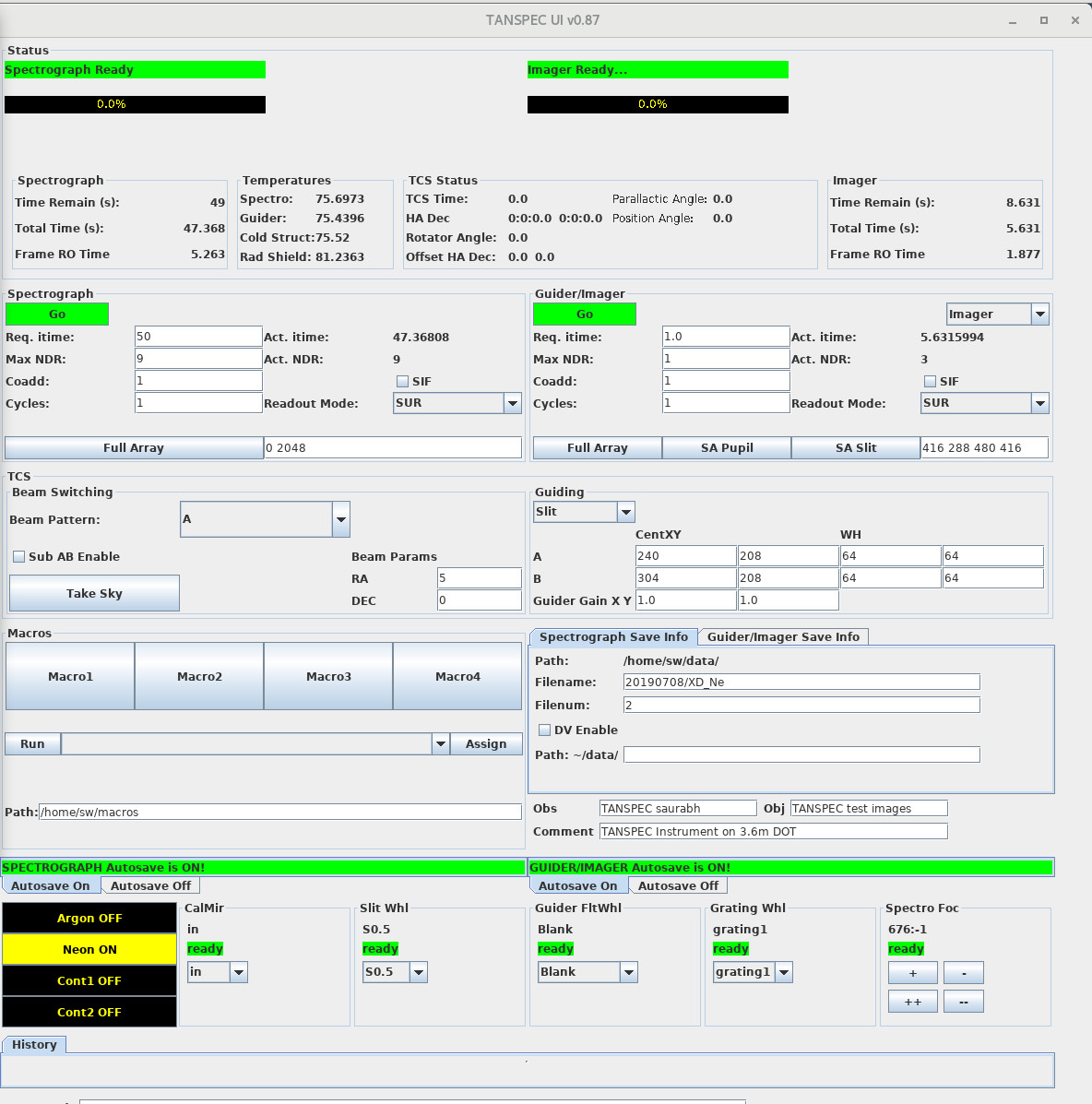}
	\caption{The TANSPEC graphical user interface (GUI). It provides control to the TANSPEC by using various buttons, icons, menus,	and macro files. }
	\label{fgui}
	\end{figure*}

	\section{Software} \label{SoftwareSection}

	The computer in the observing room hosts software to communicate and control individual components
	such as  arrays, mechanism motors, lamps, etc.
	The software is also interfaced with the Telescope Control System (TCS) and has guiding capabilities. 
	It also fetch header information from the TCS. It has Temperature Control Software for the spectrograph detector and Temperature Monitor Software
	for the optical bench and Guide Detector. All the mechanism can be controlled/powered ON through this
	Software along with capabilities to do Scripting and Macros. 
	In Fig \ref{fgui}, we show the screen-shot of the GUI of the TANSPEC
	which provides control to the TANSPEC by using various buttons, icons, menus,
	and macro files.
	
\begin{figure*}
	\centering
	\includegraphics[width=0.75\textwidth, angle=0]{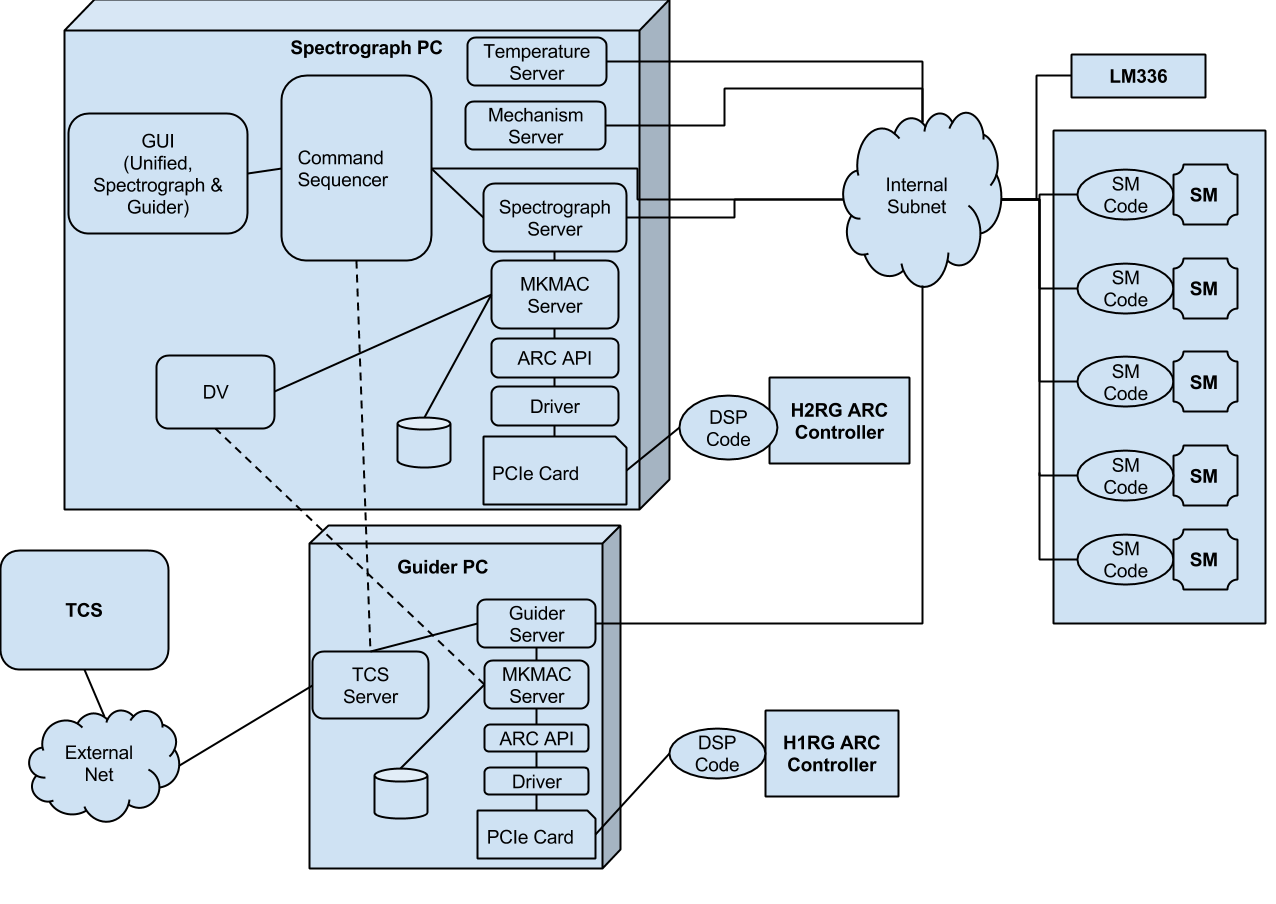}
	\caption{A representative scheme of the TANSPEC software. A PC (both for spectrograph and slit-viewer/guider) is placed in the observing room with different servers, 
	PCI cards, and software. The PC is also connected with TCS (for its feedback/control), array controllers, different mechanisms, etc. } 
	\label{fsoftware}
	\end{figure*}

	The overall structure of the TANSPEC software  (see also, \cite{2014JAI.....350006N})  is shown in Fig \ref{fsoftware}.
	The central server is the command sequencer and it
	communicates and control other servers via UNIX style sockets. The command sequencer receives the command from the GUI and sends corresponding 
	instructions to other servers (e.g., mechanism server, Array Controller Server (ACS), etc), and then, returns instrument status to the GUI. 
	The mechanism server and ACS sends commands to the, Animatics Smart-Motor via the Ethernet link, and  ARC controller via the ARC application programming interface (API), respectively.  
	After the data is fully written to the system memory, PCI board sends interrupt command to the  Central Processing Unit (CPU) and ACS generates Flexible Image Transport System (FITS) image
	from the raw data and saves it to the hard disk. It also sends the data via a socket to
	the quick look viewer software \doclink{http://irtfweb.ifa.hawaii.edu/Facility/DV/}{DV, data viewer by NASA Infrared Telescope Facility (IRTF), Hawaii, USA}. DV has a capabilities to do basic image arithmetic as well as quick measurements to check image quality.

	\section{Detectors and its performances} \label{Detectors_and_its_performance}

	Two science grade detectors, H2RG  and H1RG,  procured
	from \doclink{https://www.teledyne.com/en-us}{Teledyne, USA} are used in the spectrograph and slit-viewer, respectively. 
	Their characterization results are  provided below.
	
	\subsection{Readout mode and image generation}

\begin{figure*}
	\centering
	\includegraphics[width=0.35\textwidth, angle=0]{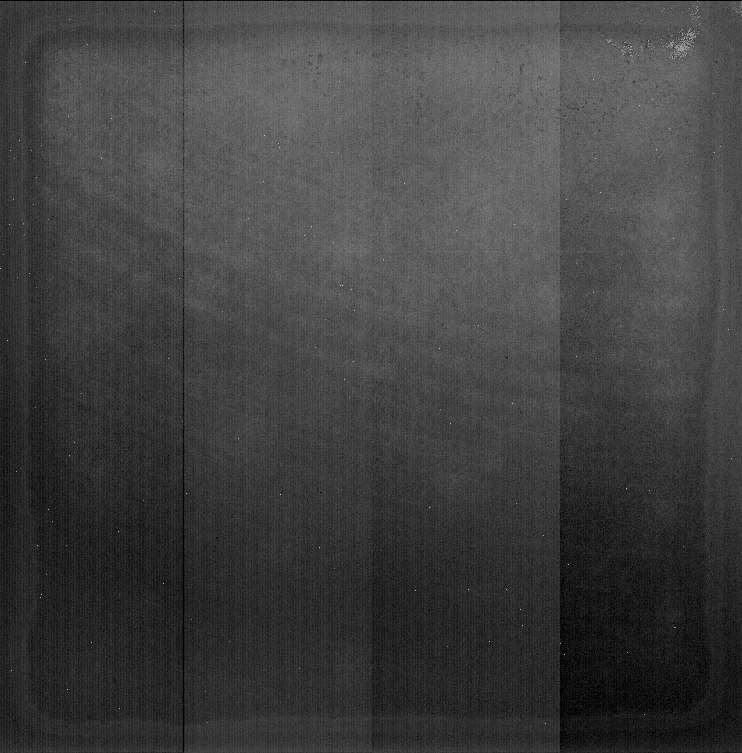}
	\includegraphics[width=0.46\textwidth, angle=0]{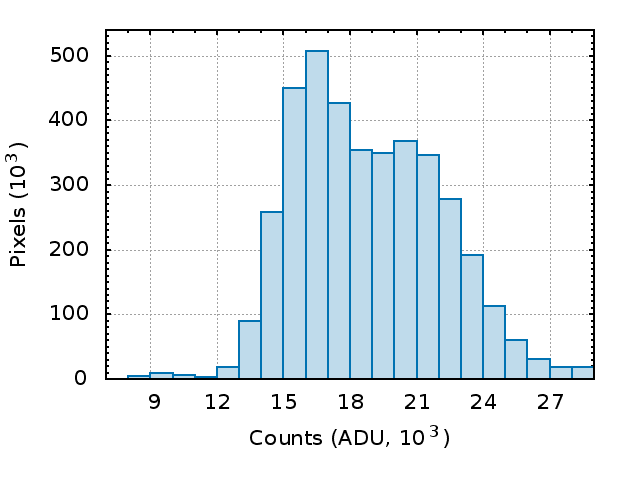}
	\includegraphics[width=0.35\textwidth, angle=0]{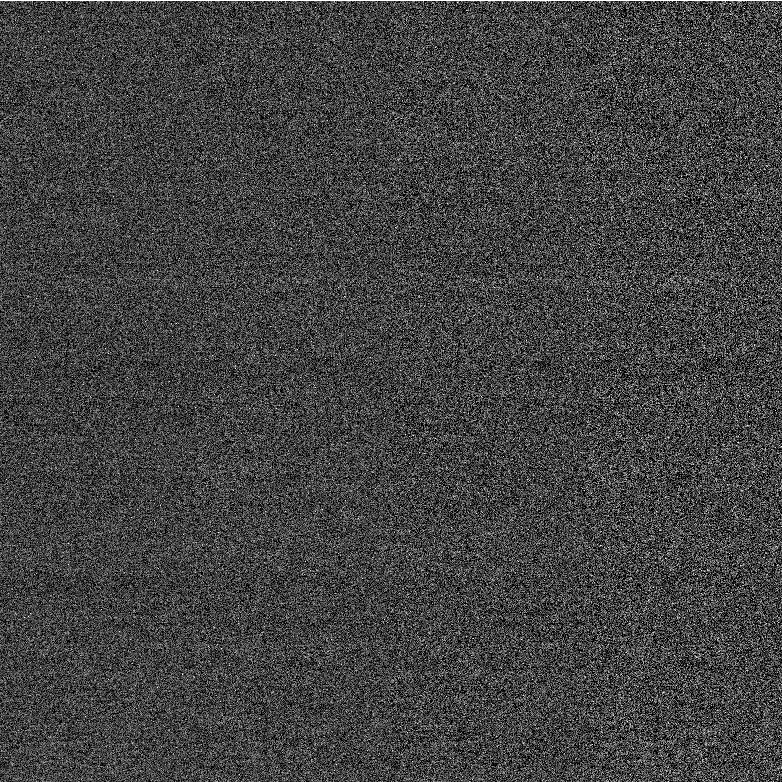}
	\includegraphics[width=0.46\textwidth, angle=0]{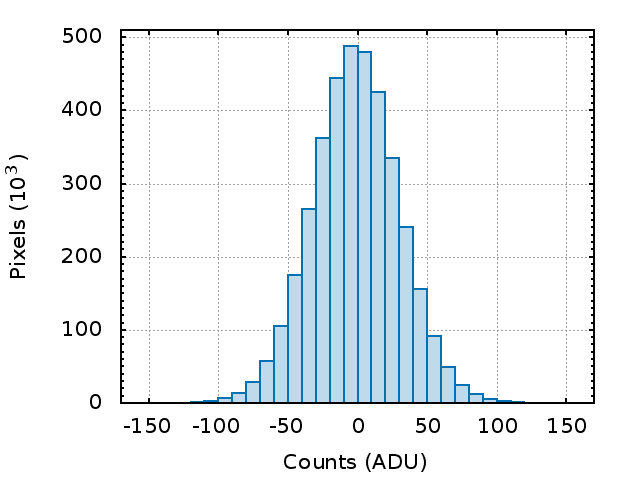}
	\caption{Upper panels: Image and histogram of the first readout (bias level) of the dark image for the H2RG array in high gain setting. Lower panels: Similar as above but for the second readout after subtracting the first readout.}
	\label{dark-image}
	\end{figure*}

	The array is continuously reset pixel by pixel when an  exposure is not being taken. When an exposure is requested the present background reset is completed and then the requested exposure is done. This reduces the after image and avoids hard saturating array and also it sets up a consistent thermal cadence and reduces the thermal anomaly effect.
	The detector can be read out in various modes, i.e. Single Read, Double Correlated Sampling and the Sample-Up-The-Ramp (SUTR).
	The SUTR is the recommended mode of readout as there will be reduction in the readout noise 
	due to the multiple sampling of the ramp values which reduces the readout noise to about 5 electrons.
	The SUTR readout technique is also useful when the pixels saturated or are hit by cosmic rays (see for details,  \cite{2014JAI.....350006N}).
	The flux rate in each pixel is calculated by fitting a line to the counts along the time axis on the SUTR
	readout data cube. 
	The H2RG array has sub-array
	capabilities in Y-axis, whereas, the H1RG has this capability in both the X and Y axes of the array. 
	There is also in-built software to guide on slit image and send corrections to TCS Software to guide on-field objects.
	In H2RG array, four vertical-strips are read out simultaneously at a rate of $\sim$5 $\mu$s per pixel and
	in total,  5.263 seconds are required to complete one full frame read out.
	Therefore, the  exposure time is  adjusted automatically to the nearest multiple of 5.263 second (non-destructive 
        readouts of the cumulative counts) and is written to the header of the image file.
	For H1RG array, it take 1.877 seconds to readout the frame and therefore, the actual exposure time is
	adjusted to the multiple of 1.877 second.
	In the top panel of Fig \ref{dark-image}, we show the image and distribution of the counts for the first readout of the H2RG array.
	As the H2RG is readout via 4 channels, we can see four strips for that in the image. A broad intensity distribution with high values as expected in the bias level of NIR detectors is seen.
	In the lower panel of Fig \ref{dark-image}, the same has been shown for the second readout by subtracting the first readout.
	The sampling clearly shows a Gaussian distribution with mean around zero and sigma of $\sim$25  Analog to Digital Unit (ADU) 
	which is equivalent to the readout noise of the H2RG array in low-gain setting. 
		
        \subsection{Gain}

\begin{figure*}
        \centering
        \includegraphics[width=0.49\textwidth, angle=0]{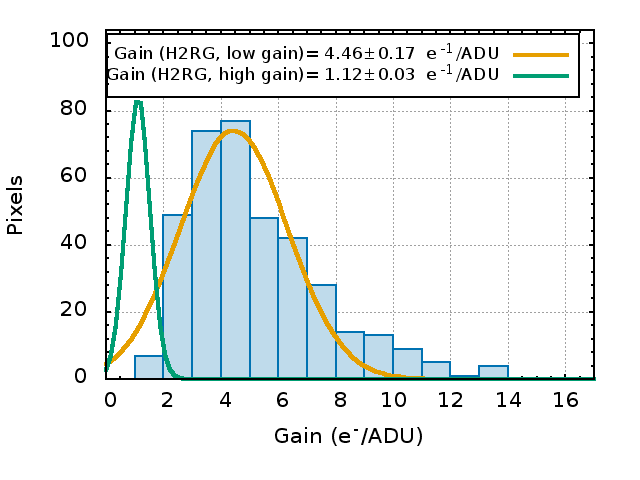}
        \includegraphics[width=0.49\textwidth, angle=0]{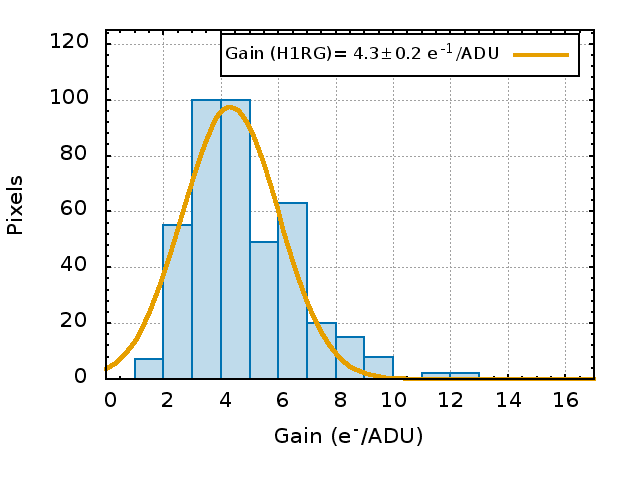}
        \caption{Distribution of Gain values for the H2RG array in low/high-gain settings (left panel) and for the H1RG array (right panel).
	The thick line is a Gaussian fitted curve on the distribution.}
        \label{gain}
        \end{figure*}

        We have taken many frames of different exposures  with the
        array exposed to the continuum lamp (for H2RG) and to the telescope enclosure with $Br\gamma$ filter (for H1RG).
	Then, for each pixel, we have plotted the counts and its variance (also know as photon transfer curve) and calculated the slop.
	The inverse of this slope will be the gain value  (see for details,  \cite{2014JAI.....350006N}).
        We have selected different regions of box size 5 $\times$ 5  pixels of the exposed area, totaling to $\sim$500 pixels, 
	of both H1RG and  H2RG arrays for the gain calculation. 
	Fig \ref{gain} shows the histograms of the gains obtained for both H2RG and H1RG arrays. 
	We have fitted a Gaussian function on these distributions and found the peak gain values as
        $4.5\pm0.2$ $e^-$/ADU  and $4.3\pm0.2$ $e^-$/ADU for the H2RG and H1RG arrays, respectively.     
	In high-gain setting of the H2RG array, we have found peak gain value of $\sim1.12\pm0.03$ $e^-$/ADU.   
        
	\subsection{Dark Current}

\begin{figure*}
	\centering
	\includegraphics[width=0.49\textwidth, angle=0]{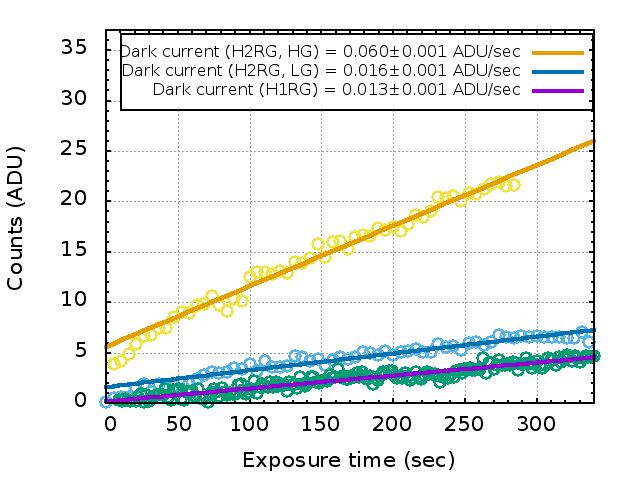}
	\includegraphics[width=0.49\textwidth, angle=0]{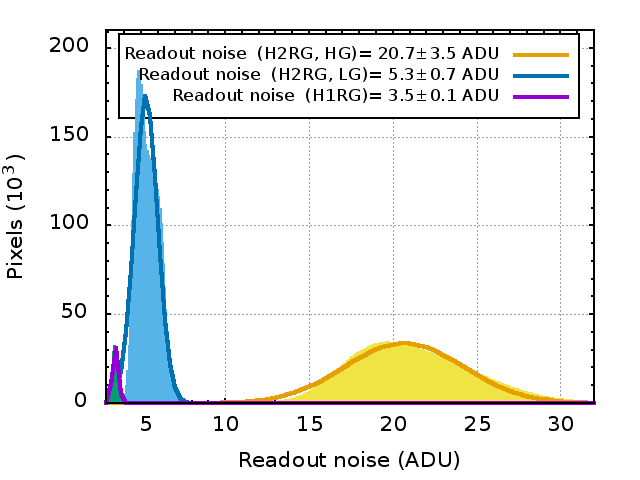}
	\caption{Dark current (left panel) and readout noise (right panel) 
	for the H2RG array in both high-gain (HG) and low-gain (LG) settings and for the H1RG array.  
	}
	\label{dark}
	\end{figure*}

	Dark current is estimated by putting cold block filter in the filter wheel (for H1RG) or the mirror in the slit wheel (for H2RG) and
	reading out the array.
	In the left panel of Fig \ref{dark}, we show the variation of the counts in the dark frame as a function of time for 
	the H2RG (for both high-gain and low-gain settings) and the H1RG arrays. A straight line fit to this distribution
	gave a dark current value of $\sim$0.016 ADU/sec and  $\sim$0.013 ADU/sec for H2RG (low-gain) and H1RG arrays, respectively.
	This multiplied by gain-values gives us an estimate of the dark current as $\sim$0.07 e$^-$/sec and  $\sim$0.06 e$^-$/sec. 
	Similarly, in the high-gain setting of the H2RG array, the dark current is $\sim$0.07 e$^-$/sec. 
	Few sets of dark frame readouts  at the beginning and end of each night would be enough for its correction. 
	
	\subsection{Readout Noise}

       	Since dark current is very small, for the readout noise calculation, we have used set of frames generated from the difference of consecutive frames
       	of a dark exposure.
       	We have calculated the standard deviation in the count values for each pixel in the above set of frames.
       	This standard deviation is actually $\sqrt{2}$ times readout noise value of each pixel (see for details, \cite{2014JAI.....350006N}).
       	The distribution of the readout noise for all pixels in the arrays are shown in the right panel of Fig \ref{dark}. 
	The readout noise for the H2RG (low-gain) and H1RG arrays is found be $5.3\pm0.7$ and $3.5\pm0.1$ ADUs, respectively.
	In the high-gain setting of the H2RG, the readout noise is estimated as $20.7\pm3.5$ ADUs.
	Here, it is worthwhile to mention that the above readout noise values are for a single readout. The readout noise
    is a function of the number of readouts and will decrease with increase in the number of readouts, e.g., in case of H2RG, the readout noise will be 1.2 ADUs for 284 readouts (or 5.3 e$^-$, low gain setting). For more details please see \cite{2019PASP..131e5001W}.

	\subsection{Linear range and Saturation levels }

\begin{figure*}
	\centering
	\includegraphics[width=0.49\textwidth, angle=0]{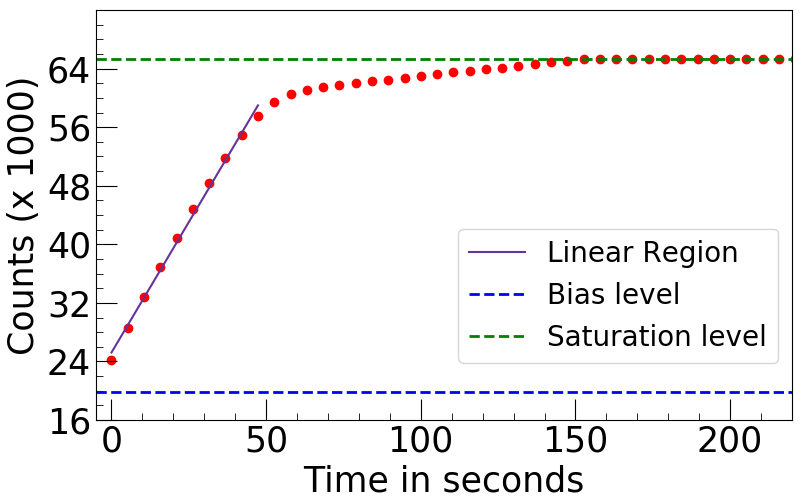}
	\includegraphics[width=0.49\textwidth, angle=0]{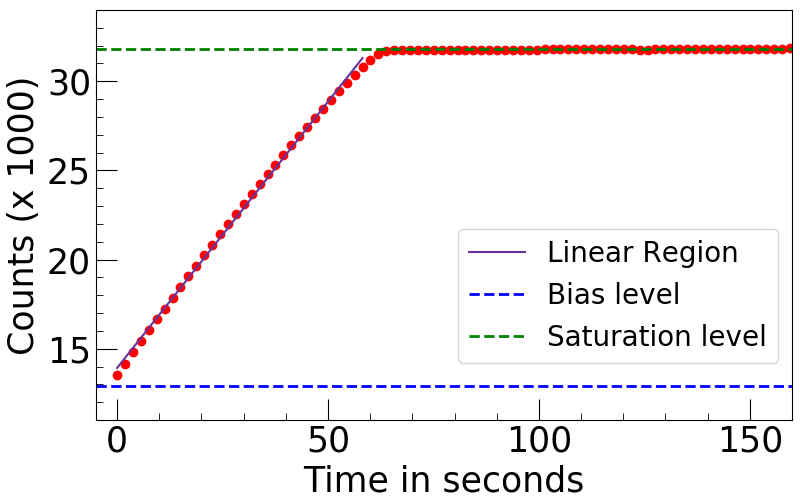}
        \caption{Linear range  and the saturation values for the H2RG (left panel) and H1RG (right panel) arrays.}
	\label{saturation}
	\end{figure*}

	Using the SUTR readout data of continuum lamps (for H2RG array) and Argon lamp with Br$\gamma$ filter (for H1RG),
	we have estimated the median saturation level of these arrays as $\sim$28000 ADU ($\sim$65000 ADU for high gain) and $\sim$32000 ADU, respectively, for low gain setting.
	In Fig \ref{saturation}, we show the variation of the median intensity value as a function of exposure time for these arrays for low gain setting.
	We can clearly see the saturation levels along with the non-linearity in the curves near these  limits. 
		We have estimated the median value of the upper limit for the linear regime of H2RG and H1RG arrays as 26000 ADU (57000 ADU for high gain) and 30000 ADU, respectively.
	As the median bias level for the H2RG and H1RG arrays are at $\sim$20000 ADU (20000 ADU for high gain) and $\sim$13000 ADU, the effective (useful) well depths 
	for these arrays are found to be at the level of $\sim$6000 ADU ($\sim$37000 ADU for high gain) and $\sim$17000 ADU, respectively, in low gain setting.
	Here, it is worthwhile to mention that very bright spectral lines or stars can result in residual images that can contaminate the darks.

	\subsection{Bad, Hot and Cold pixels}

	The pixels which deviate more than 8$\sigma$ from the median value in
	the image  which  is generated by dividing two flats (taken in low  and high incident flux), are defined as bad pixels. 
	The percentage of the bad pixels in the H1RG  array is 0.03\%, most of which are along
	the edges. 
	For the H2RG array, which is used for spectroscopy, the different orders observed in continuum lamps were used to detect 
	bad pixels which are negligible in numbers.
 
	The pixels having counts more than 8$\sigma$ above the median value in a dark readout of 100 sec are defined as hot pixels. The fraction
	of hot pixels for the H1RG and H2RG arrays was found to be 0.05\% and 0.03\%, respectively. They are distributed randomly over the arrays. 
	Similarly, cold pixels (pixels which have counts less than 8$\sigma$ below
	median value in a dark readout of 100 seconds) fraction was found to be negligible for both the arrays. 
	
	The main parameters of the H2RG and H1RG arrays used in the TANSPEC instrument are listed in Table \ref{param}.

        \begin{table*}
        \centering
	\scriptsize
        \caption{\label{param} Main parameters of the  detectors used in the TANSPEC instrument.}
        \begin{tabular}{@{}cccc@{}}
        \hline
        Parameter                                       &    H2RG (high-gain)   &H2RG (low-gain) &  H1RG       \\
        \hline
        Number of pixels				&	$2048\times2048$&	$2048\times2048$	&	$1024\times1024$\\	
	Pixel size ($\mu$m)				&	18		&	18		&	18			\\
	Gain            (e$^-$/ADU)               	&       1.1             &       4.5     &   4.3          \\
        Readout noise    (ADU)                		&       $20.7\pm3.5$   &       $5.3\pm0.7$      &       $3.5\pm0.1$       \\
        Dark Current (ADU/sec)           	        &       $0.060\pm0.001$     &    $0.016\pm0.001$           & $0.013\pm0.001$             \\
	Saturation (ADU)				&	65000		&	28000&		32000\\
	Use full well-depth (ADU)			&	37000		&	6000&		17000\\
	Readout time (sec)				&    5.263		&   5.263			& 1.877\\
	Readout channels 				&	4       	&	4       	& 1       	\\
	Sampling					&Single read, Up the Ramp, Double Correlated Sampling,	&same& same	\\
							&Non Destructive Reads (NDRs)&	-	&\\
	Sub-array capability				&	In Y -axis	&	In Y-axis	&	In both X and Y axes\\
	\hline
        \end{tabular}\\
        \end{table*}

	\section{Performance on the 3.6-m DOT} \label{Performance_on_the_3.6-m DOT}

	Ground testing of the TANSPEC was carried out during the first week of April 2019 and subsequently it was mounted on the 3.6-m DOT. The TANSPEC was then used for on-sky commissioning tests during April - May, 2019. We present initial results in the following sub-sections.

	\begin{figure*}
	\centering
	\includegraphics[width=0.40\textwidth, angle=0]{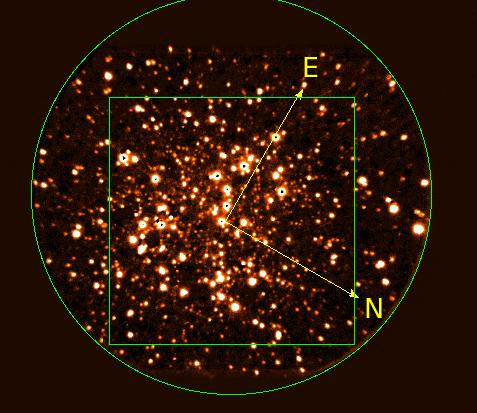}
	\caption{Image of a Globular cluster M53 taken in the $K_s$ band using the slit-viewer of the TANSPEC on 3.6-m DOT on 20 April 2019. 
	The sub-array of the H1RG array is used to image a circular FOV of sky having radius of 49
        arcsec (shown as a green circle) through a mirror in the slit wheel and a filter in the filter
        wheel. The image quality of the FOV is optimized for the 60$^\prime$$^\prime \times$ 60$^\prime$$^\prime$ FOV (shown as a green box).
	The FOV of the slit viewer is rotated by 120 degrees in clockwise direction which can be aligned by off-setting  rotator of the telescope by 120 degrees.}
	\label{m53}
	\end{figure*}

	\subsection{Observing procedure}

 	The detailed procedure for observing through the TANSPEC is provided in \doclink{https://aries.res.in/sites/default/files/files/3.6-DOT/Tanspec-Observation-Manual.pdf}{the TANSPEC observation manual.}
	Briefly, the observer selects the required slit and spectroscopy mode (XD mode or prism mode) through slit wheel and grating wheel, respectively. 
	A desired filter can be selected through the filter wheel and the target source is then imaged in the slit viewer/guider.
	Afterwards, the target source is placed in the center of the slit by giving off-sets to the telescope.
	Guiding can also be done with the slit viewer through any filter (optical to NIR) 
	on an object in its FOV.
        The magnitude limit for auto-guiding is roughly about J $\sim$10 mag in an exposure of 10 s. 
	Telescope's visible guider can also be used to guide stars.
	Once guiding is started, long integration 
	of several hundred seconds can be done on the faint sources. 
	In order to correct the spectra for the sky lines, telluric standards should be observed near to the target source at about similar airmass.
	Calibration frames such as arc-lamps, dark, flats are usually taken at standard star position by  executing a calibration macro.
	All updates and information about the TANSPEC instrument are provided at \doclink{https://aries.res.in/facilities/astronomical-telescopes/360cm-telescope/Instruments}{ARIES 3.6-m DOT instrument page.}
	The initial version of the data reduction pipeline, \doclink{https://github.com/astrosupriyo/pyTANSPEC}{pyTANSPEC}, which supports the XD
	mode data reduction for slits of width 0.5 and 1.0 arcsec is also released.

	\begin{figure*}
	\centering
	\includegraphics[width=0.35\textwidth, angle=0]{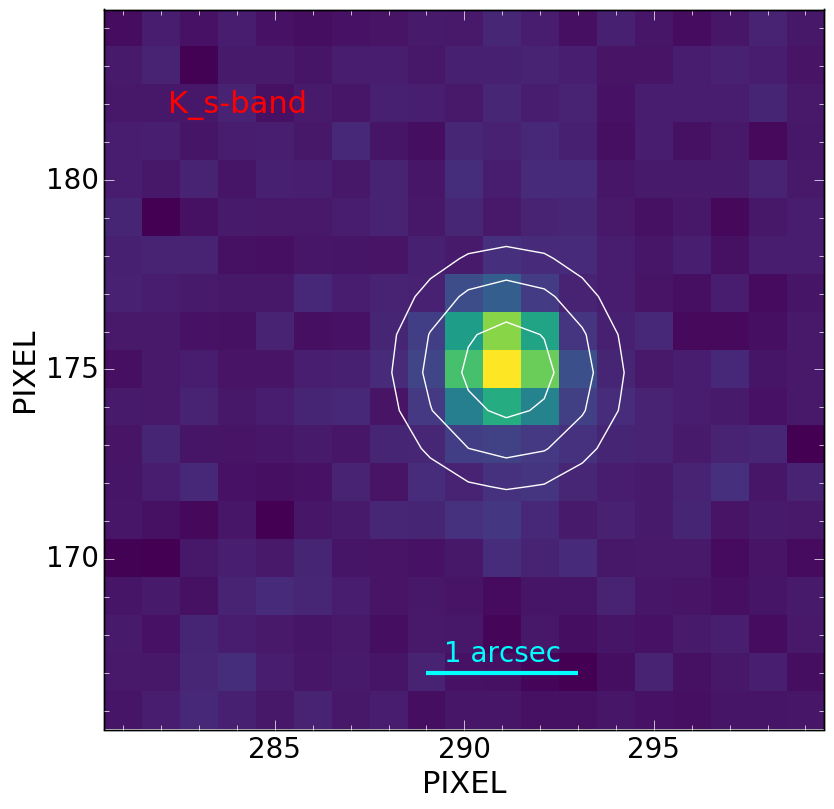}
	\includegraphics[width=0.45\textwidth, angle=0]{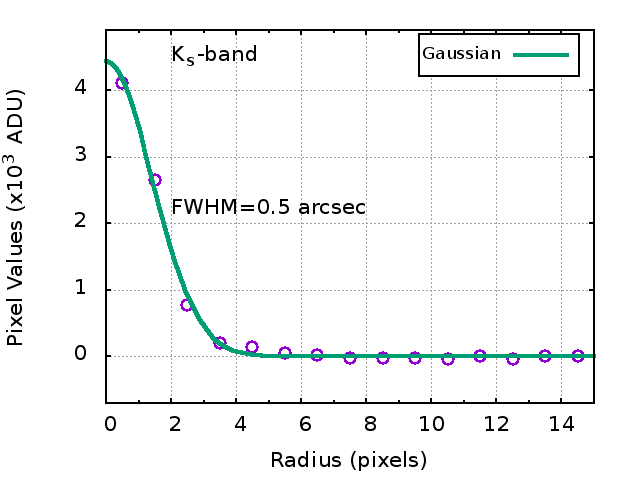}
	\includegraphics[width=0.35\textwidth, angle=0]{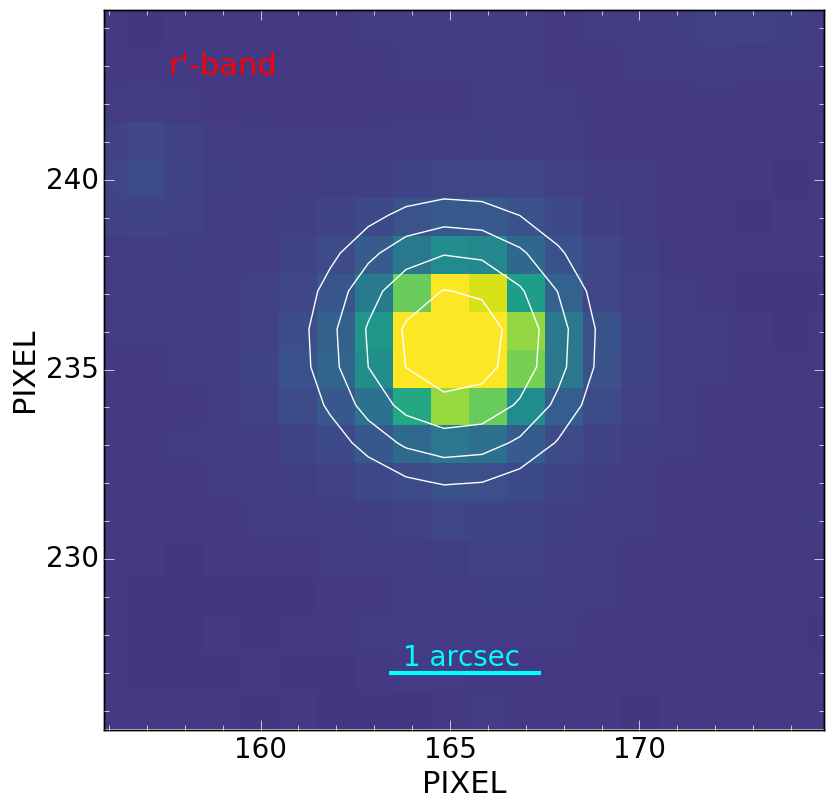}
	\includegraphics[width=0.45\textwidth, angle=0]{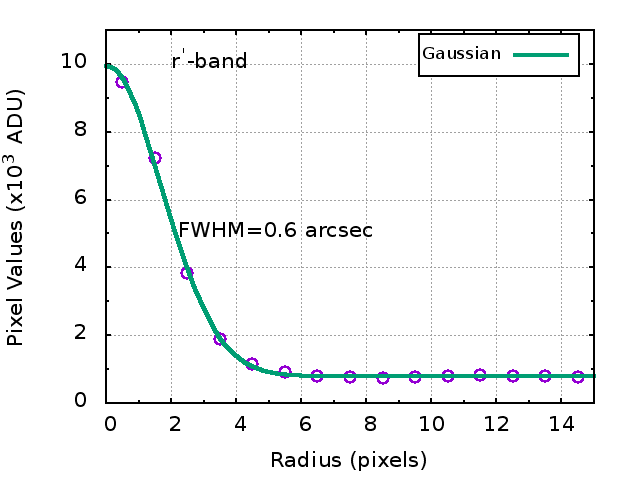}
	\caption{Shape and radial profile of stars in $K_s$ (upper panels) and $r^\prime $ (lower panels) bands using the slit-viewer on the TANSPEC.
	The full width at half maximum (FWHM) of the stellar profile is 0.5 arcsec and 0.6 arcsec in $K_s$ and $r^\prime $ bands, respectively.}
	\label{iq}
	\end{figure*}

	\subsection{Imaging: Slit-viewer}

	By using the slit-viewer in the TANSPEC mounted on the 3.6-m DOT, we  can image celestial objects on the H1RG array. 

	\subsubsection{Field-of-view, Plate Scale and Image Quality}
	
    Image of a Globular cluster M53 taken in the $K_s$ band using the slit-viewer of the TANSPEC on 3.6-m DOT
    is shown in  Fig \ref{m53}.
	The sub-array of the H1RG array is used to image a circular FOV of sky having radius of 49
	arcsec (shown as a green circle in the Fig \ref{m53}) through a mirror in the slit wheel and a filter in the filter
	wheel. The image quality is optimized for the 60$^\prime$$^\prime \times$ 60$^\prime$$^\prime$ FOV (please refer Fig \ref{m53}). 
        Astrometry of the image of the astronomical objects obtained by using the slit-viewer was done by using the 
	\doclink{http://star-www.dur.ac.uk/~pdraper/gaia/gaia.html}{Graphical Astronomy and Image Analysis Tool} with a rms noise of the order of $\sim$0.25 arcsec.
	The plate scale was estimated as 0.245$^\prime$$^\prime$/pixel. The FOV the slit-viewer is rotated
	by 120 degrees in clockwise direction which can be aligned in the celestial north-south direction by off-setting rotator of the telescope by 120 degrees.

	 An example of typical stellar profiles in  $K_s$ and $r^\prime $ bands using the slit-viewer on the TANSPEC are shown in Fig \ref{iq}. We have observed several regions in photometric nights to check for the variation of the full width at half maximum (FWHM) in different directions.
	The size (FWHM) of the point spread function (PSF) of the stellar images is found out be
	around $\sim$0.45-0.60 arcsec in the $K_s$-band (2.2 $\mu$m) and  $\sim$ 0.6-0.7 arcsec in the $r^\prime$-band (0.61 $\mu$m). 
	This is in accordance to the \doclink{https://www.astro.auth.gr/~seeing-gr/seeing_gr_files/theory/node17.html}{wavelength dependence of the atmospheric seeing.} 
	The images are also found to be circular in shape (ellipticity = 0.02 - 0.2).
	There is no direction dependent variation in the PSF of the stellar images.

	\subsubsection{Color equations}

        We took NIR observations of a field near W51 star-forming complex ($\alpha$$_{2000}$ =19$^{h}$25$^{m}$40$^{s}$,
	$\delta$$_{2000}$ = +15$^\circ$07$^\prime$46$^{\prime\prime}$) with  four frames of 10 sec exposure in each of the 5-point dither pattern 
	in $J,~H,$ and $K_s$ filters. 
        We also did observation of an open cluster Be 20 ($\alpha$$_{2000}$ =05$^{h}$32$^{m}$36$^{s}$,
        $\delta$$_{2000}$ = +00$^\circ$11$^\prime$19$^{\prime\prime}$) in optical $r^\prime $ and $i^\prime $ bands with 12 frames of 5 minutes exposures, i.e., a total of one hour exposure in $r^\prime $ and $i^\prime $ filters.
        Dark frames and sky flats were also taken during the observations.
        The sky frames in NIR filters ($JHK_s$) were generated by median combining the dithered frames.
        The basic image processing such as dark/sky subtraction and flat fielding was done using tasks
        available within Image Reduction and Analysis Facility (IRAF)\footnote{IRAF is distributed by the National Optical Astronomy Observatory,
        which is operated by the Association of Universities for Research in Astronomy (AURA)
        under cooperative agreement with the National Science Foundation.}.
        Instrumental magnitudes were obtained using the {\sc daophot} package.
        As the target region was  crowded, we have carried out the PSF photometry to get the magnitudes of the stars.
        The obtained color correction equations for NIR filters ($JHK_s$) of the TANSPEC on the 3.6-m DOT for calibrating to 
	Two Micron All Sky Survey (2MASS) magnitudes are as follows (see also, \cite{2022ApJ...926...68G}).

	\begin{equation}
	(J-H) = (0.91\pm0.07)\times(j-h) + 0.05\pm0.08
	\end{equation}

	\begin{equation}
	(H-K) = (1.01\pm0.07)\times(h-k) + 0.51\pm0.05
	\end{equation}

	\begin{equation}
	(J-K) = (1.17\pm0.05)\times(j-k) -0.70\pm0.12
	\end{equation}

	\begin{equation}
	(J-j) = (0.004\pm0.04)\times(J-H) -1.21\pm0.04
	\end{equation}

        where, $JHK$ are the standard magnitudes of the
        stars taken from the 2MASS catalog \cite{2003yCat.2246....0C} and $jhk$ are the present instrumental magnitudes of the same stars normalized per sec exposure time.

        The color correction equations obtained for optical filters ($r^\prime $ and $i^\prime $) of the TANSPEC on the 3.6-m DOT for calibrating to the Panoramic Survey Telescope and Rapid Response System (Pan-STARRS1 or PS1 \cite{2016arXiv161205560C}) are obtained as follows.

	\begin{equation}
	(r_{P1} - i_{P1})= (0.99\pm0.03)\times(r-i)  -0.06\pm0.01
	\end{equation}
	\begin{equation}
	(r_{P1} - r) =  (0.08\pm0.06)\times(r_{P1} - i_{P1})  -6.14\pm0.02
	\end{equation}
	\begin{equation}
	(i_{P1} - i) =  (-0.01\pm0.01)\times(r_{P1} - i_{P1})  -6.05\pm0.02
	\end{equation}

        where, $r_{P1}$ and $i_{P1}$ denote the standard magnitudes of the
        stars taken from the \doclink{http://catalogs.mast.stsci.edu/}{PS1 data release 1 catalog}, and  $r$ and $i$ are the present instrumental magnitudes of the similar stars normalized per sec exposure time.

	\subsubsection{Sensitivity}

 	To estimate the photometric sensitivity, we took deep NIR observations of an open cluster Teutsch 76 ($\alpha$$_{2000}$ =22$^{h}$28$^{m}$47$^{s}$,
       	$\delta$$_{2000}$ = +61$^\circ$37$^\prime$55$^{\prime\prime}$) with  five frames of 20 sec exposure in each of a 7-point dither pattern
       	in $J,~H,$ and $K_s$ filters. We took 5 sets of these observations which gives a total exposure time of $\sim$1 hour in $J,~H,$ and $K_s$ filters.
       	For optical bands, we have used the deep data of an open cluster Be 20 with an hour exposure each in $r^\prime $ and $i^\prime $ filters, as discussed in the previous section.
       	Individual cleaned images were average combined to give the final image on which we have done photometry as per the procedure mentioned in the previous section.
       	The FWHMs of the stellar profile in these final images were around $\sim$0.8 arcsec (NIR bands) and $\sim$1 arcsec (optical bands), respectively.
       	The observations were carried out at an approximate elevation of 55-45 degrees. The humidity was around 40-60 \%, and temperature ranges from 4 to 12 degrees Celsius.

\begin{figure*}
	\centering
	\includegraphics[width=0.32\textwidth, angle=0]{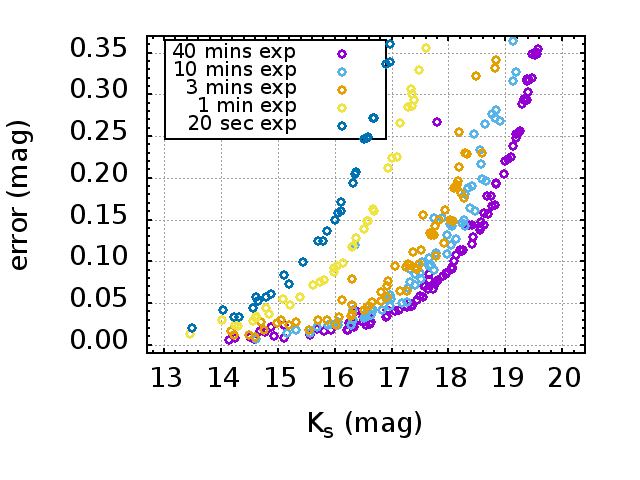}
	\includegraphics[width=0.32\textwidth, angle=0]{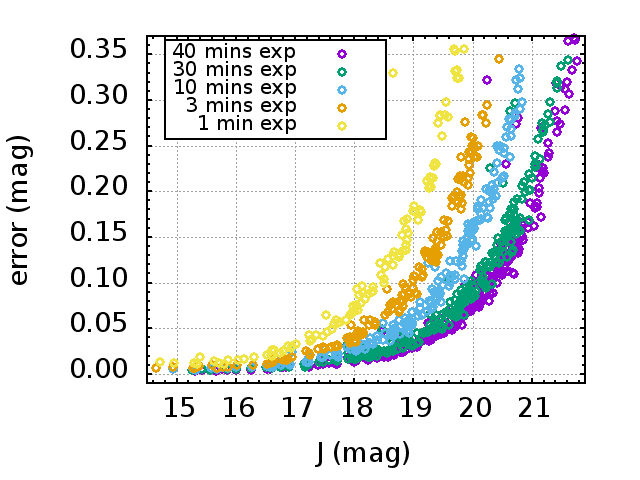}
	\includegraphics[width=0.32\textwidth, angle=0]{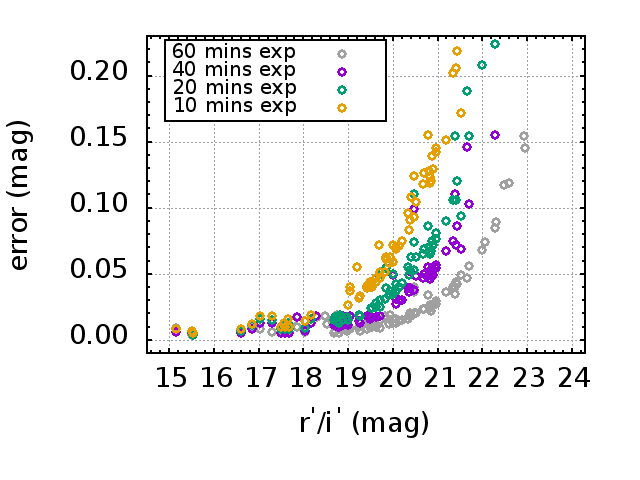}
	\includegraphics[width=0.32\textwidth, angle=0]{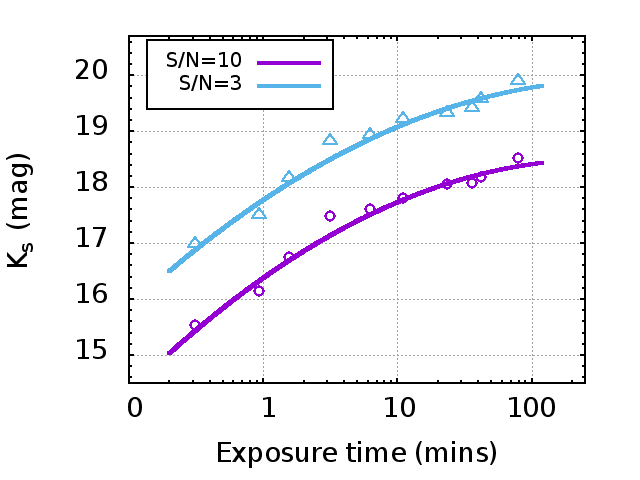}
	\includegraphics[width=0.32\textwidth, angle=0]{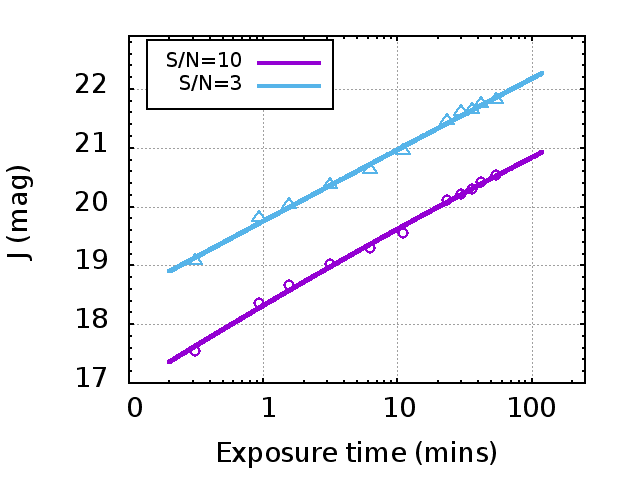}
	\includegraphics[width=0.32\textwidth, angle=0]{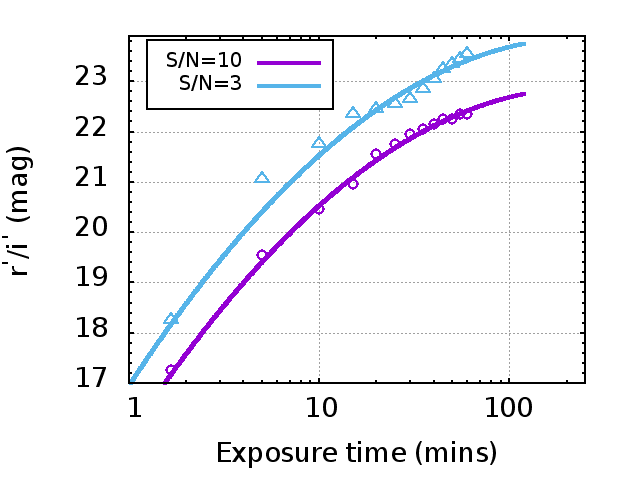}
	\caption{Sensitivity curves for the slit-viewer in the TANSPEC instrument.
	Top panels: Error as a function of magnitude in $K_s$, $J$, and $r^\prime /i^\prime $ bands for different exposure times. Bottom panels: $K_s$, $J$, and $r^\prime /i^\prime $ magnitudes as a function of exposure time for SNR = 10 and 3. The solid curves are the fitted function on the data points shown by open circles/triangles.}
	\label{sensehk}
	\end{figure*}

       	The top panel of Fig \ref{sensehk} shows the magnitude versus  
       	error plot for different exposures for the stars in the cluster field in the IR and optical filters.
       	Using these plots, we have generated the exposure time versus magnitude graphs in different filters for signal to noise ratio (SNR) = 10 and SNR = 3 (detection limit)
       	and are shown in the bottom panels of the Fig \ref{sensehk}.
	The curved lines are the polynomials fitted to the data points generated by the smoothed spline function on the observed data points.
       	These curves are generated with an M1 reflectively value of 90 \%.
	Clearly, the TANSPEC has a sensitivity to detect faint source of $K_s\sim19.7$ mag or $r^\prime \sim23.7$ (with an SNR of 3) in two hours of exposure.
	An exposure time calculator based on these curves is under development.

	\subsection{Spectroscopy: XD mode and prism mode}

	\subsubsection{Wavelength range}

       The calibration unit of the TANSPEC has Argon and Neon lamps  for wavelength calibration and two continuum lamps for flat fielding
       the spectra taken in both XD and prism modes. The Argon lamp covers mostly
       the NIR part of the spectra whereas the Neon lamp covers mostly the optical part. 
       Two continuum lamps, one at high voltage and another at low voltage, are used to cover all the orders of the spectra. 
	The aperture on the integrating sphere is fine-tuned in such a way that
       the spectra do not get saturated in both modes of the TANSPEC.

\begin{figure*}
	\centering
	{\hspace{.1cm}
	\includegraphics[width=0.40\textwidth, angle=0]{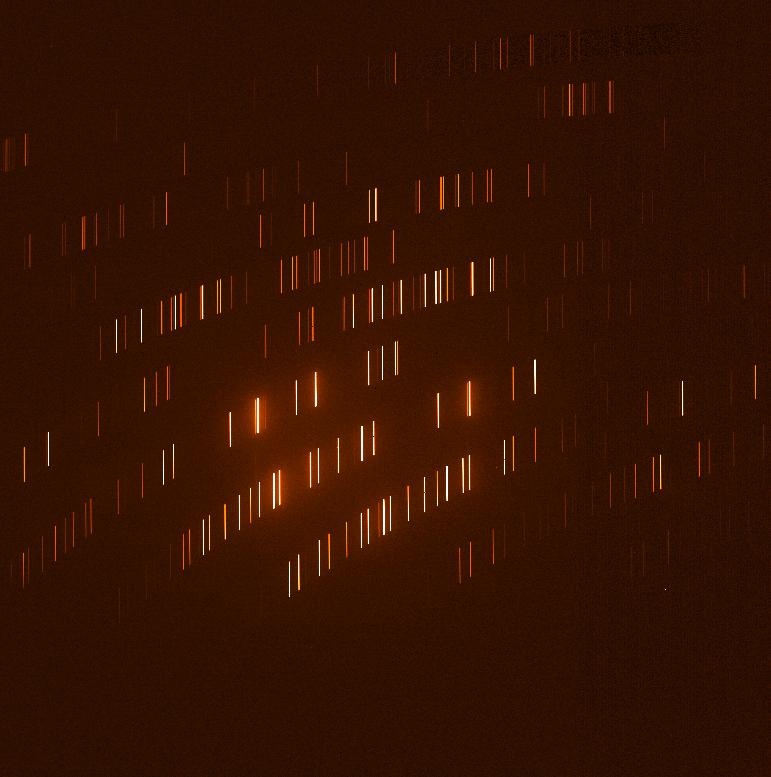}
	\hspace{0.2cm}
	\includegraphics[width=0.40\textwidth, angle=0]{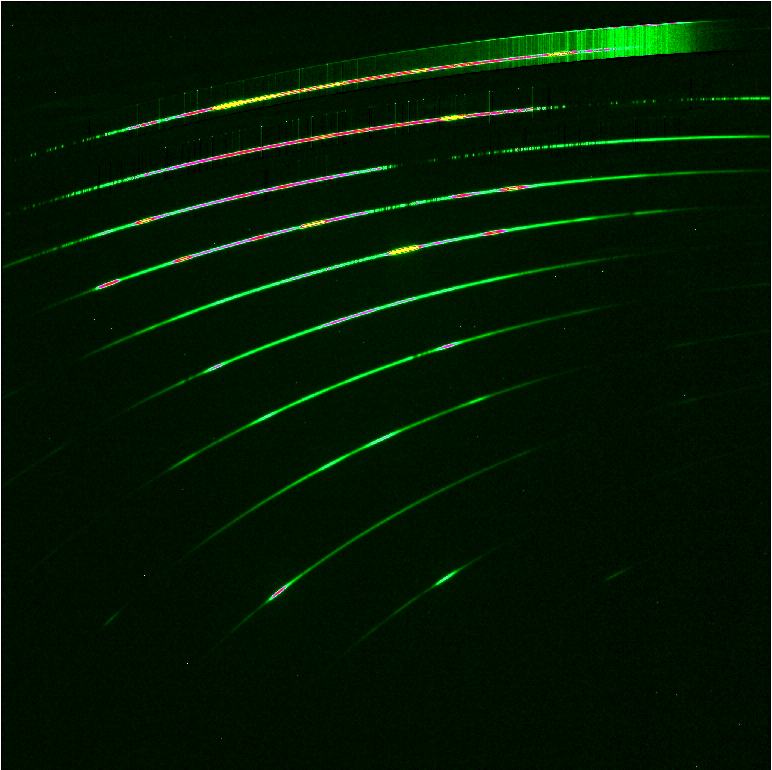}}
	{\hspace{.3cm}
	\includegraphics[width=0.41\textwidth, angle=0]{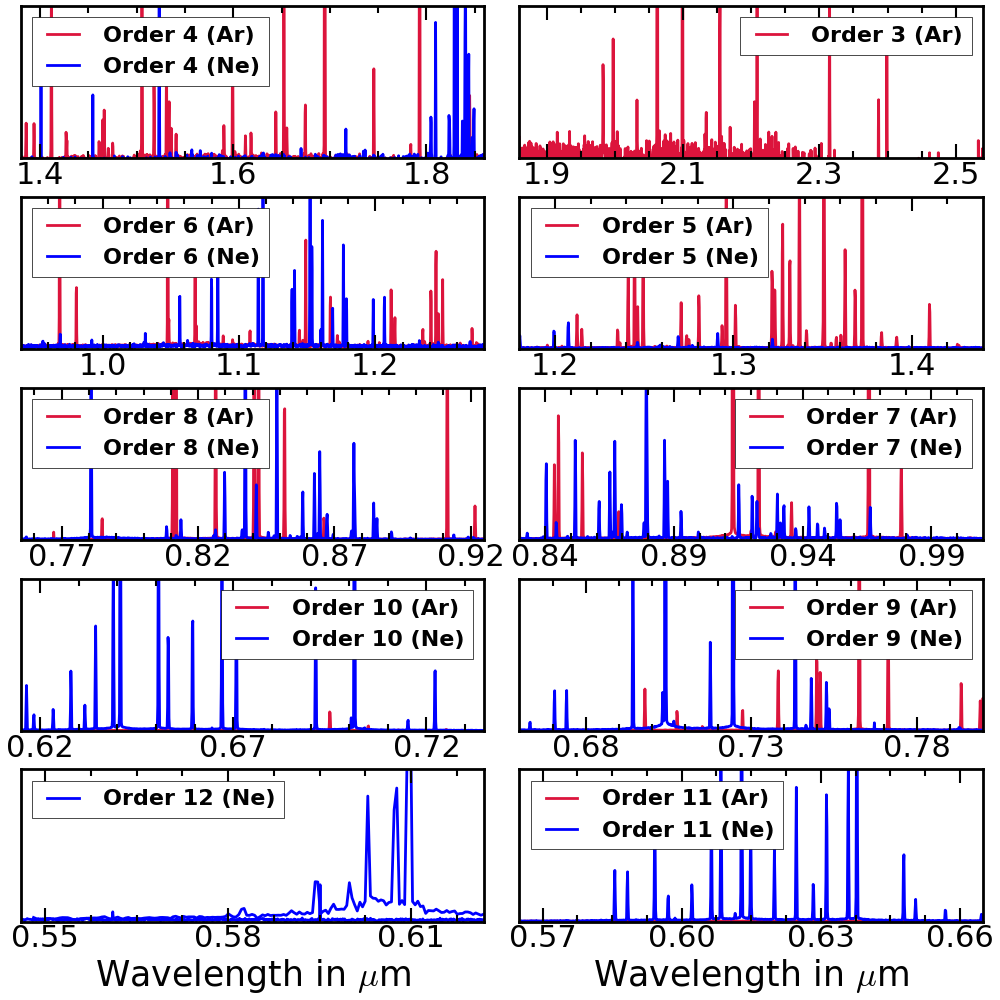}
	\hspace{0.2cm}
	\includegraphics[width=0.41\textwidth, angle=0]{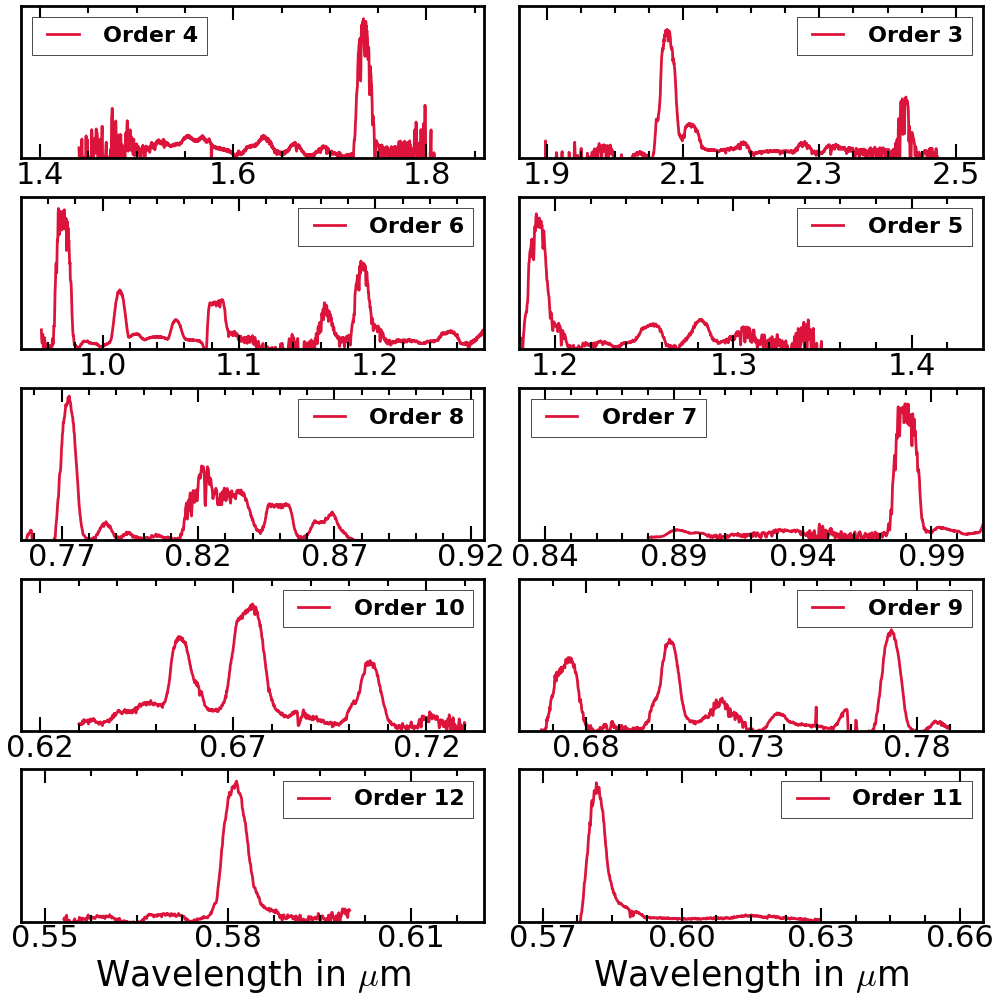}}
	\caption{Images of the Neon lamp spectra (top left) and the spectra of an WR star, `HD 16523' (top right) taken through TANSPEC in the XD mode. 
	The wavelength calibrated spectra in different orders of the XD mode of the TANSPEC for Neon/Argon lamps (bottom left) and an WR star (bottom right).}
	\label{images-spectra-xd}
	\end{figure*}

       In the top panels of Fig \ref{images-spectra-xd}, we show the images of the Neon lamp (left panel) and the stellar spectra of an Wolf Rayet (WR) star `HD 16523' (right panel)
       in the XD mode of the TANSPEC. These images were taken with 0.5 arcsec slit having 20 arcsec length.
       Exposure times for Argon/Neon lamps and an WR star ($K\sim7.9$ mag) were 120/100 sec and 450 sec, respectively.
       All the orders starting from 3$^{rd}$ (1.86 - 2.54 $\mu$m) to 12$^{th}$ (0.55- 0.57 $\mu$m) are detected in both the lamps and stellar spectra.
       The spectra of an WR star clearly show the broad emission line features in almost all orders.
       The sky lines and thermal background at longer wavelengths ($\lambda\gtrsim2.3 ~\mu$m) are also seen in the stellar spectra.
       The sky lines are mostly OH emission and can be subtracted by the spectra of a telluric standard.
       The thermal background can be due to several reasons, e.g., thermal emission from the telescope and dome structure, and can add noise in that part of the spectra.
       As can be seen from the images, all orders in the lamp spectra have sufficient lines that can be used to wavelength calibrate the stellar spectra.
       The lower panel of Fig \ref{images-spectra-xd} shows the extracted wavelength calibrated spectra for both lamp (left panel) and stellar (right panel) for all the orders.
       They also show the good coverage of lines required for wavelength calibration as well as various spectral features of an WR star.
       Wavelength calibration was done using the {\sc identify} task in IRAF with an accuracy of 0.1$\mu$m and the reciprocal linear dispersion was $\sim$1-4 \AA/pixel.

\begin{figure*}
	\centering
	
	\includegraphics[width=0.47\textwidth, angle=0]{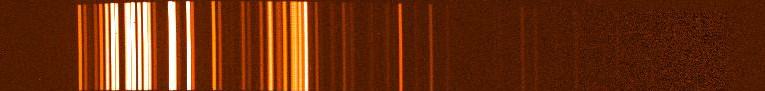}
	\hspace{.15cm}
    	\includegraphics[width=0.47\textwidth, angle=0]{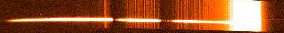}
	{\vspace{.2cm}
	\includegraphics[width=0.47\textwidth, angle=0]{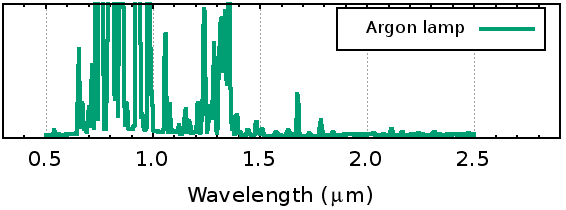}
    	\hspace{.15cm}
	\includegraphics[width=0.47\textwidth, angle=0]{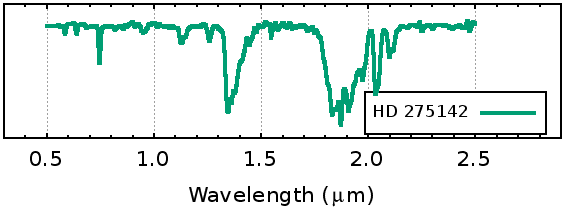}}
	\caption{Images of the Argon Lamp spectra (top left) and  F-type star `HD 275142' spectra (top right) taken through the TANSPEC in the prism mode. 
	The wavelength calibrated spectra in the prism mode of the TANSPEC for Argon lamp (bottom left) and `HD 275142' star (bottom right).}
	\label{images-spectra-lr}
	\end{figure*}

       Similarly, in the top panels of Fig \ref{images-spectra-lr}, we show the images of the Neon lamp (left panel) and stellar spectra of a F-type star `HD 275142' (right panel)
       in the prism mode of the TANSPEC. These images were also taken using the 0.5 arcsec slit having 20 arcsec length.
       Exposure times for Argon/Neon lamps and HD 275142 ($K\sim9.7$ mag) were 3.5 sec and 150 sec, respectively.
       The lower panel of Fig \ref{images-spectra-lr} shows the extracted wavelength calibrated spectra for both lamp (left panel) and stellar (right panel) in the prism mode.
       From Figs \ref{images-spectra-xd} and \ref{images-spectra-lr}, we can also see the full wavelength coverage of the instrument from 0.55 to 2.5 $\mu$m.
       Sky lines and thermal background are also evident in the spectra.
       No stellar features are seen at some wavelength regimes because of the atmospheric absorption due to H$_2$O and OH bands.
       Wavelength calibration was done with an accuracy of 0.4 $\mu$m and  the reciprocal linear dispersion was $\sim$1-4 \AA/pixel.

	\subsubsection{Spectral Resolution}

\begin{figure*}
	\centering
	\includegraphics[width=0.48\textwidth, angle=0]{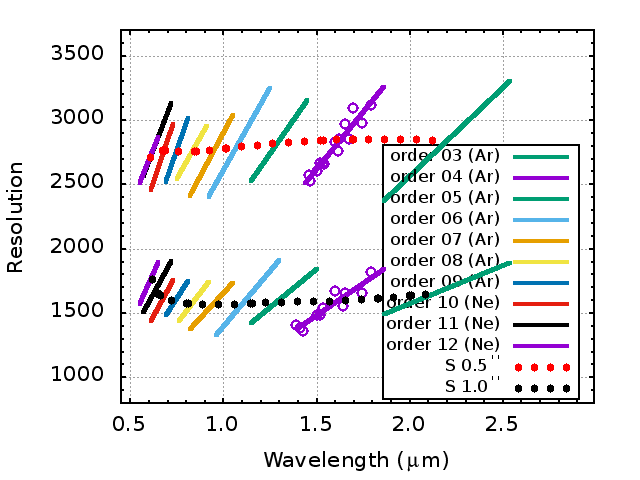}
	\includegraphics[width=0.48\textwidth, angle=0]{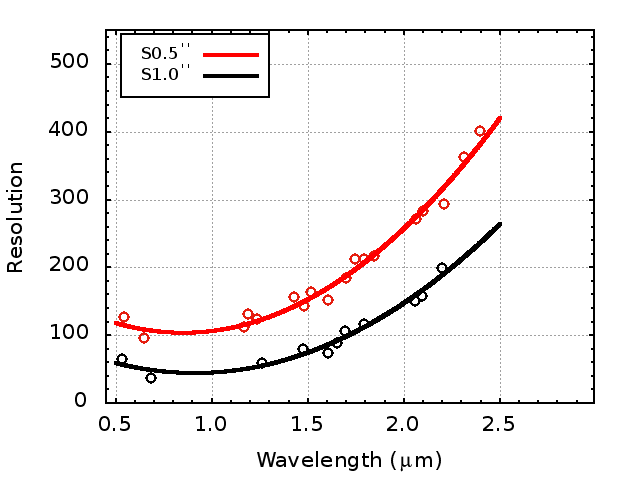}
	\caption{Spectral resolution as a function of wavelength in the XD mode (for different orders, left panel) and the prism mode (right panel) of the TANSPEC for 0.5 arcsec and 1 arcsec wide slits.
	The solid curves are fitted functions on the data points shown as open circles (for clarity, we have shown data points only for the 4th order in the XD mode).}
	\label{resolution}
	\end{figure*}

	The effective spectral resolving power of the spectra in each order was estimated by finding 
	FWHMs of Argon/Neon lamp lines through the slits of 0.5 arcsec and 1.0 arcsec widths. 
	Fig \ref{resolution} shows the spectral
	resolving power ($\lambda/\Delta \lambda$) as a function of wavelength for 0.5 arcsec and 1.0 arcsec slits as obtained from the Argon/Neon lines  
	in the XD mode (left panel) and the prism mode (right panel) of the TANSPEC. 
	The median resolution in the XD mode is  estimated as  2750 and 1600 for 0.5 arcsec and 1.0 arcsec slits, respectively.
	For the prism mode, the resolution varies from 100 (50) to 400 (250) for the 0.5 (1.0) arcsec slit across different regions of the spectra, respectively.
	
	\subsection{Throughput}

	For the efficient throughput, the entire optics needed to work efficiently across the entire 0.55-2.5 $\mu$m range of the instrument. 
	The theoretical estimates of the throughput of the TANSPEC instrument in different modes at different wavelengths are given in Table \ref{tp}.
	These numbers are estimated for the best optics performance of the system including the
        telescope mirrors, instrument mirrors/lens/prism/grating/substrate and quantum efficiencies of the arrays at different wavelengths. 
	However, these estimates  can be treated as upper limits as
	they do not include the atmospheric transmission and the slit losses.
	As expected, there is a decrease in throughput at shorter wavelengths ($r^\prime $ and $i^\prime $ bands) due to
	the combination of decreasing quantum efficiency (QE) of the array (including its own single-layer anti-reflection coating) 
	and decreasing efficiency of the Broadband Anti-Reflection (BBAR) coatings on the transmissive optics.
	The peak throughput of the XD mode is about $\sim$30\%. On the other hand, the peak throughput
	of the prism mode ( $\sim$50\%) is far more than that of the XD mode
	The imaging mode peak throughput is estimated as $\sim$60\%.

        \begin{table*}
        \centering
        \caption{\label{tp}Estimate of the throughput (in \%) of the TANSPEC  Instrument for best optics performance including the 
	telescope mirrors, instrument mirrors/lens/prism/grating/substrate and the quantum efficiencies of the detector arrays at different wavelengths.}
        \begin{tabular}{@{}c|ccccccc@{}}
        \hline
	Bands ($\mu$m)		&	$r^\prime $(0.66)&	$i^\prime $(0.81)&	$Z$(0.9)&	$Y$(1.02)&	$J$(1.22)	&	$H$(1.63)	&	$K_s$(2.19)\\
        \hline
	Imaging mode   				&	  42	&	  40	&	  51	&	  55	&	  58	&	  60	&	  63\\
	XD mode					&	  22	&	  20	&	  26	&	  28	&	  29	&	  31	&	  33\\
	Prism mode				&	  37	&	  33	&	  42	&	  45	&	  48	&	  52	&	  36\\
        \hline
        \end{tabular}
        \end{table*}

	The throughput of the entire system in the imaging mode was also measured from the total flux of standard stars 
	in an area of 4 times of the FWHM of their profile.  The imaging mode throughput percentage is $30\pm5$\%, $45\pm5$\%,
	$44\pm5$\% and $35\pm5$\% for the $r^\prime $, $J$, $H$ and $K_s$ bands, respectively. 
	The lower $K_s$ band throughput is may be due to the higher background at these wavelengths, i.e. thermal background and atmospheric lines.
	These numbers also includes other parameters such as QE of the detector, instrument/telescope optics and the atmospheric turbulences. 
	The throughput in spectroscopy mode is very difficult to measure as it involves placing the star at
	the exact center of the slit every time. 
	However, the throughput in the prism and XD modes was measured on an A0V star through 1 arcsec wide slit.
	We used the $J$ band region of the spectrum to calculate the throughput. The theoretical $J$ band flux for
\setcounter{footnote}{0}
	the star was calculated using Pogson's relation\footnote{M=2.512*log(F) + C} where the zero point of the magnitude
	is taken from literature \cite{2003AJ....126.1090C}. This value was converted to photon counts by
	dividing it with the energy corresponding to the $J$ band wavelength. The observed photon counts
	were converted to units of counts/m$^2$/s by dividing it with the effective collecting area and the total
	integration time. The ratio of the observed values to the theoretical values are obtained as 21\% and 37\%, which
	are the throughput values of the spectrograph in the XD and prism modes, respectively.
	As expected, the measured throughput values are lower than the theoretical estimates due to the atmospheric transmission and the 
	current reflectivity of the primary mirror of the 3.6-m DOT, i.e., 45-60\% in optical bands.
	Slit losses can also contribute to the low throughput values and a wider slit, e.g., 4${^\prime}{^\prime}$, can help to minimize those losses.

	\subsubsection{Sensitivity}

\begin{figure*}
	\centering
	\includegraphics[width=0.32\textwidth, angle=0]{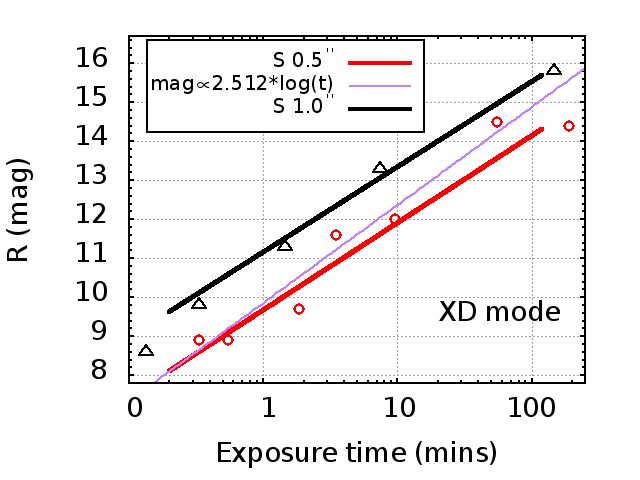}
	\includegraphics[width=0.32\textwidth, angle=0]{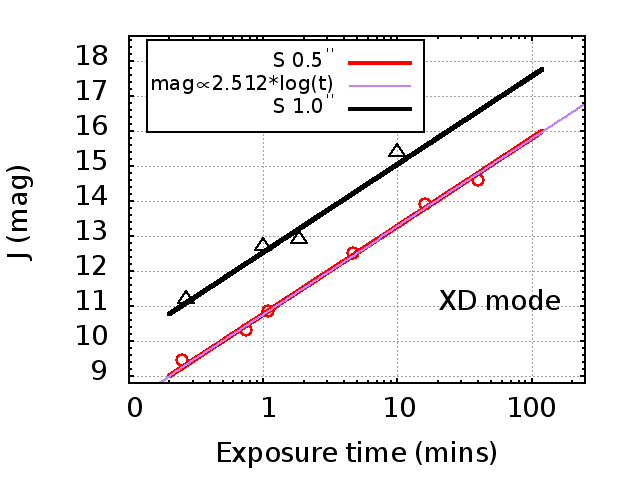}
	\includegraphics[width=0.32\textwidth, angle=0]{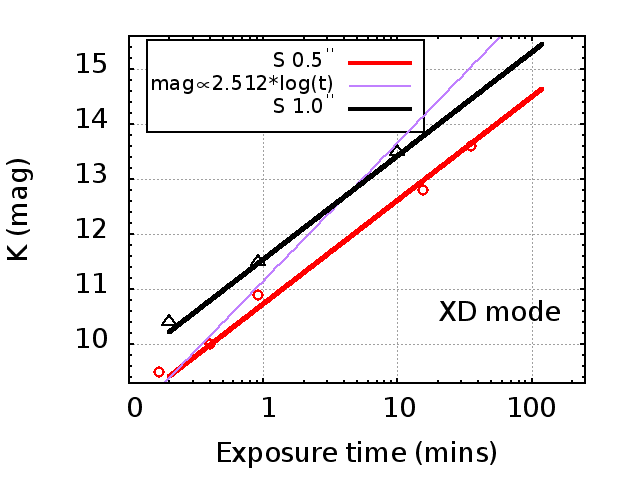}
	\includegraphics[width=0.32\textwidth, angle=0]{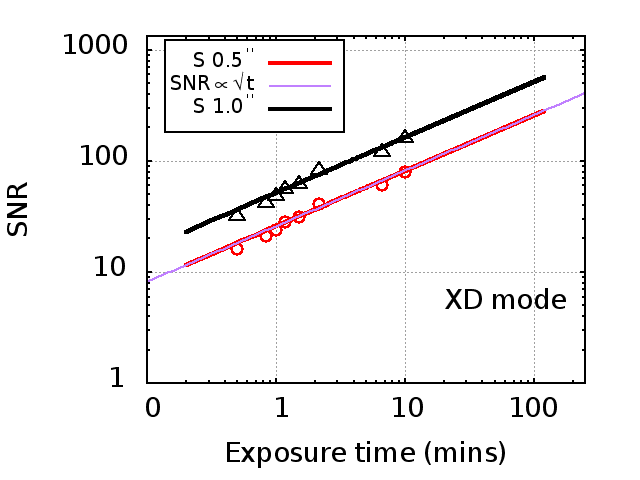}
	\includegraphics[width=0.32\textwidth, angle=0]{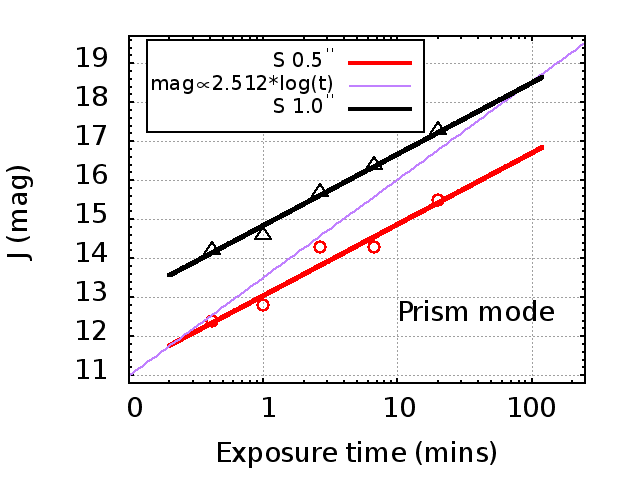}
	\includegraphics[width=0.32\textwidth, angle=0]{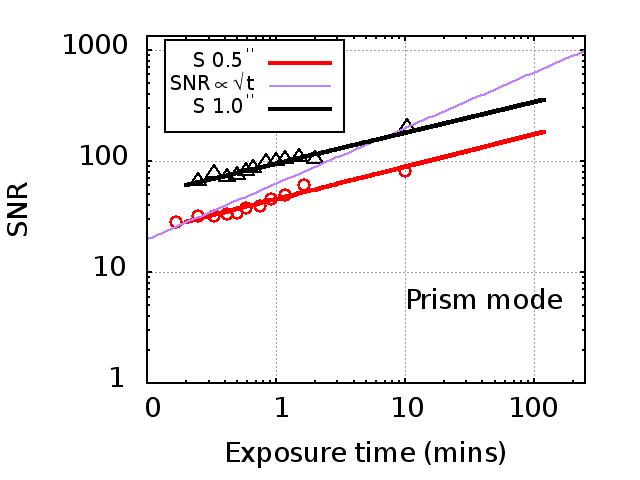}
	\caption{Sensitivity curves for the spectroscopic mode of the TANSPEC on the 3.6-m DOT. 
	Thick lines are the magnitudes (for SNR = 10) and SNR versus exposure time  curves for the XD mode and the prism mode of the 
	TANSPEC for both 0.5 arcsec (red color) and 1.0 arcsec (black color) wide slits.
	Magenta lines are the  scaling relationships (SNR$\propto \sqrt{t}$, mag$\propto$2.512*log(t)) normalized at short exposure time for the 0.5 arcsec slit.
	 The thick  curves are the fitted function on the data points shown by open circles/triangles.
	}
	\label{exp-xd}
	\end{figure*}

	Sensitivity in spectroscopic mode is estimated by taking the spectra of A0V  stars having a range of magnitudes.
	During observations, the stellar FWHM, relative humidity, temperature and the reflectivity of the M1 mirror of the 3.6-m DOT were 0.6-1.0 arcsec, $<$ 60\%, 5-15 degree Celsius and $\sim$50\%, respectively. Fig \ref{exp-xd} shows various estimates of exposure time required for a continuum SNR of 10 in different bands/orders of the XD mode as well as for the prism mode of the TANSPEC for the 0.5 arcsec and 1.0 arcsec wide slits. The SNR is quite sensitive to the FWHM of the star profile as well as the accuracy in centering of the star
	inside the slit. Therefore, the 1.0 arcsec slit provides better sensitivity than the 0.5 arcsec slit. We have found that the spectra of stars up to 15 mag ($K_s$), 17.5 mag ($J$), and 15.5 mag ($R$) can be observed in 2 hours of exposure with a 1.0 arcsec slit (with an SNR of 10) under typical night conditions. The SNR is also calculated from the extracted spectra of a $J$=10 mag A0V star in 0.5 arcsec and 1.0 arcsec slits and its variation with exposure time is also plotted in Fig \ref{exp-xd} for the XD and prism modes of the TANSPEC. Deviation of the SNR from the $\sqrt{t}$ line is possible when there are other sources of noise contributing on top of the photon noise. This can happen when counts are less, and the read noise is comparable to photon noise. The SNR at NIR bands are also prone to high background noises (thermal, atmospheric lines etc.) and can contribute to this deviation.

	\begin{figure*}
	\centering
	\includegraphics[width=0.54\textwidth, angle=0]{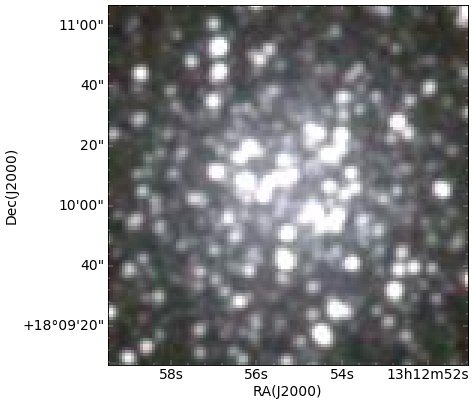}
	\includegraphics[width=0.425\textwidth, angle=0]{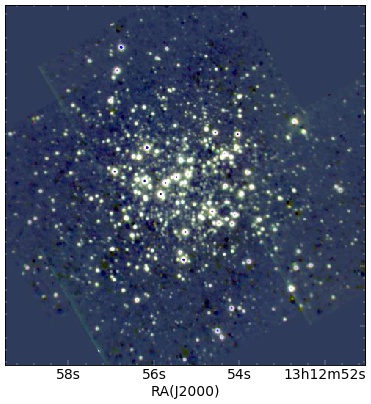}
	\caption{Comparison of the color-composite images of the $\sim2^\prime \times 2^\prime$ FOV of the central 
	region of M53 globular star cluster generated by using the $K_s$ (red), $H$ (green) and $J$ (blue) bands images from
	 2MASS (left panel) and TANSPEC (right panel).}
	\label{img}
	\end{figure*}
	
	\begin{figure*}
	\centering
	\includegraphics[width=0.407\textwidth, angle=0]{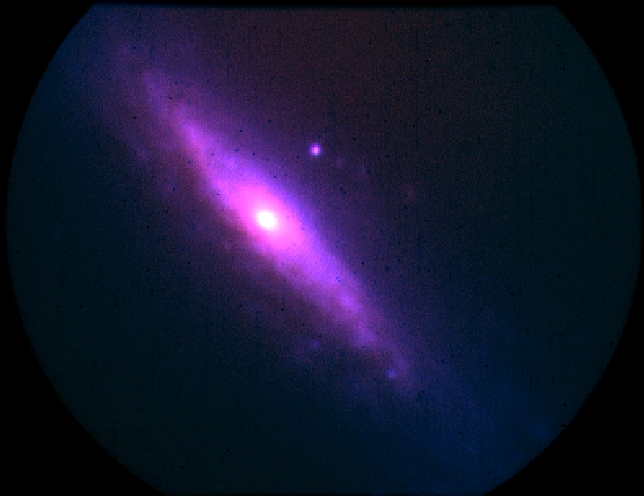}
	\includegraphics[width=0.40\textwidth, angle=0]{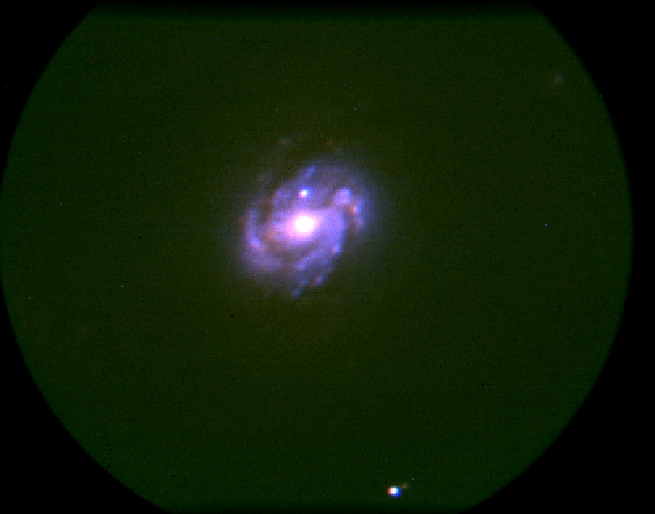}
	\includegraphics[width=0.41\textwidth, angle=0]{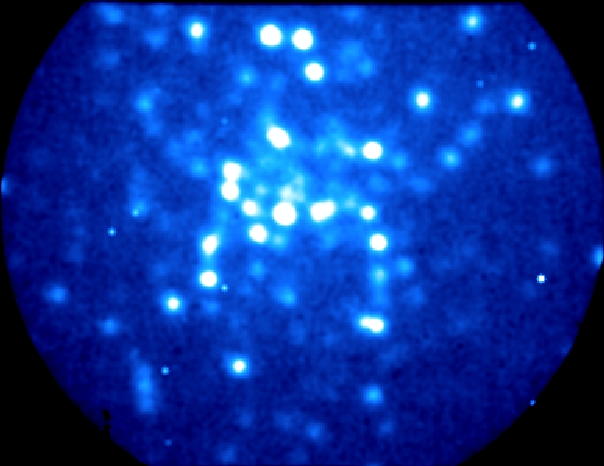}
	\includegraphics[width=0.40\textwidth, angle=0]{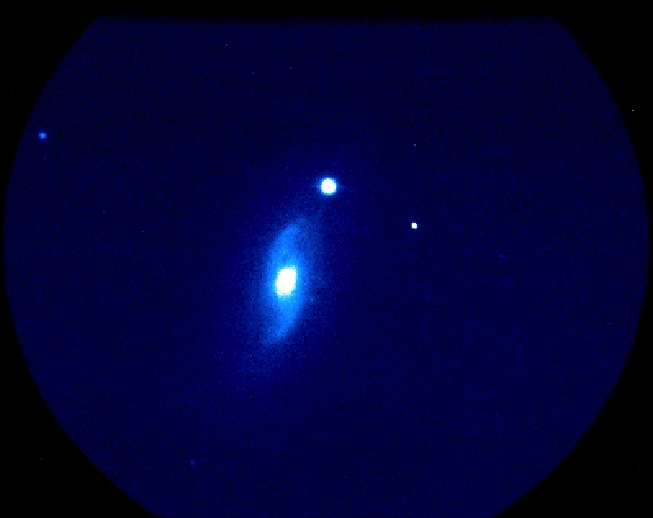}
	\caption{Color-composite images of galaxies NGC 4666 (top-left) and 
	 NGC 4328 (top-right) using $J$ (blue), $H$ (green) and $K_s$ (red) images taken through the slit-viewer of the TANSPEC. 
	$J$ band images of a globular star cluster `Palomar 2' (bottom-left) and  a Seyfert galaxy `MCG+09-16-013' (bottom-right) are also shown.}
	\label{sample-img}
	\end{figure*}

	\subsection{Example Images and Spectra}

\begin{figure*}
	\centering
	\includegraphics[width=0.95\textwidth, angle=0]{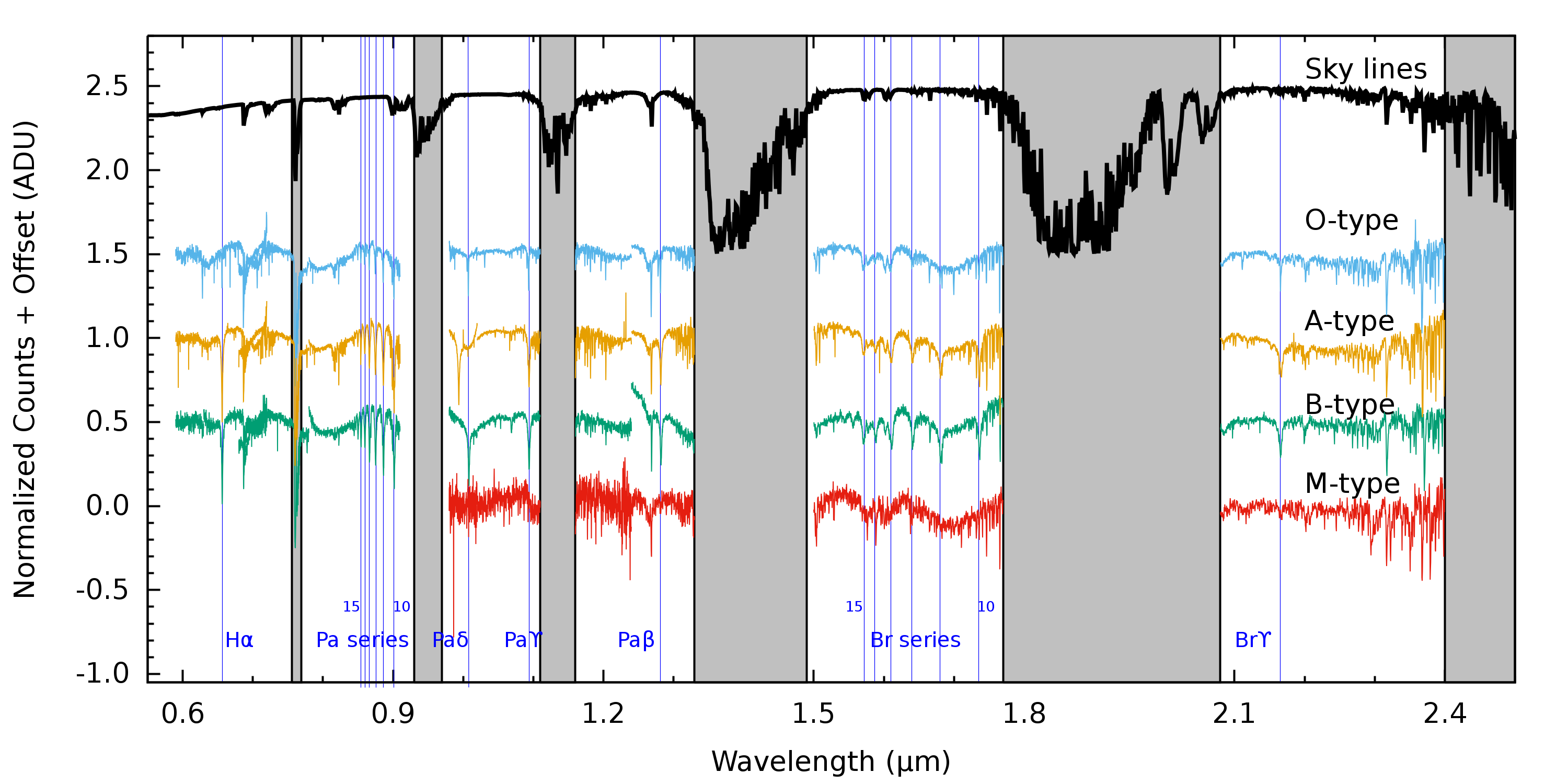}
	\includegraphics[width=0.95\textwidth, angle=0]{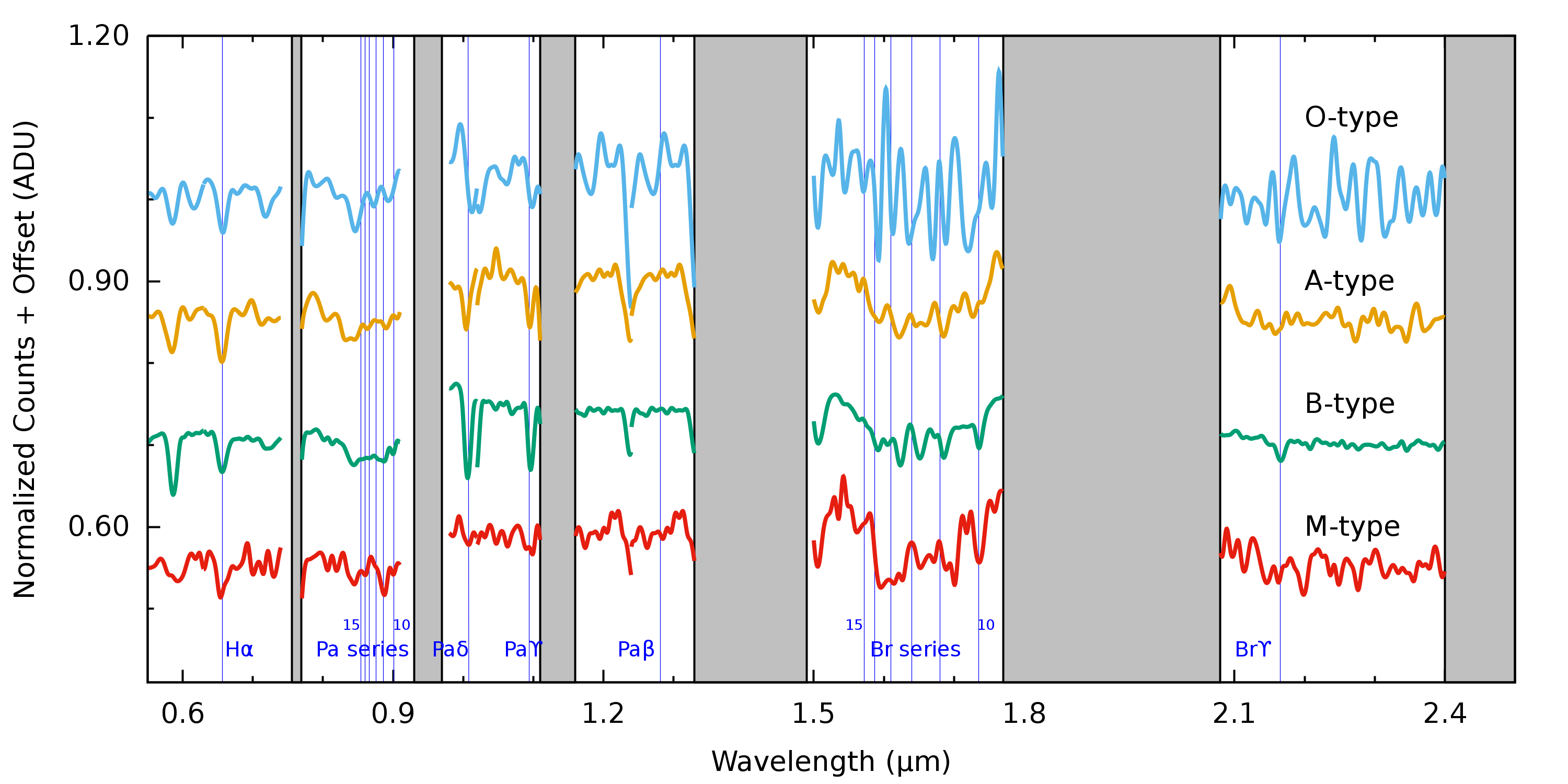}
	\caption{Sample spectra of stars of different spectral types in the XD mode (top panel) and the prism mode (bottom panel) of the TANSPEC taken with 0.5 arcsec slit. 
	The shaded regions are the regions where atmospheric features dominate.
	Various detected spectral lines are also marked.}
	\label{sample-xd-m}
	\end{figure*}

	During the commissioning nights of the TANSPEC instrument,  we have observed various
	astrophysical sources in the spectroscopy as well as imaging modes.
	We have done observations of M53 globular cluster using the slit-viewer in the TANSPEC on the
	3.6-m DOT. As the FOV of the slit-viewer is small ($\sim60^{\prime \prime} \times60^{\prime \prime} $), we have done observations of M53 in five
	pointings in $J$, $H$ and $K_s$-band filters with 25 Non Destructive Reads (NDRs). The readout time of each NDR is 1.877 secs. In Fig \ref{img}, we show
	the comparison of the mosaiced image of the slit-viewer with that of 2MASS survey. We can easily see a better
	resolved and deeper image of the TANSPEC.
	Fig \ref{sample-img} shows color-composite images created from the $J$, $H$ and $K_s$ images and $J$ band images of some 
	of the sources observed with the TANSPEC. Fig \ref{sample-xd-m}  shows the optical-NIR spectra obtained from stars with different spectral types in both the XD and prism modes of the TANSPEC.
        The black curve is the \doclink{http://twiki.cis.rit.edu/twiki/bin/view/Main/MaunaKeaTo100kmAtmosphericTransmissions}{atmospheric transmission at Mauna Kea.}
	The shaded regions are thus the regions where atmospheric features dominate the spectra and hence, we have not shown those wavelength ranges in our extracted spectra.
	We can easily identify various spectral features (marked as blue vertical lines) in both the spectra.

        \begin{table*}
        \centering
        \caption{\label{sens} Main parameters of the TANSPEC Instrument.}
        \begin{tabular}{@{}cccc@{}}
        \hline
        Parameter                                       &    XD mode            & Prism mode       & Imaging       \\
        \hline
        Wavelength coverage ($\mu$m)                    & 0.55 - 2.5              &0.55 - 2.5       &0.55 - 2.5       \\
        Resolution$^a$ ($\lambda\over{\Delta\lambda}$)      & 2750                  & $\sim$100-400 & $\sim$10      \\
        Plate scale (arcsec/pixel)                      & 0.25                  & 0.25          &0.25           \\
        Throughput (\%, at $J$ band)                    & 30                    &48             &58     \\
        Sensitivity$^b$ (mag)                           &                       &               &               \\
                $       r^\prime $                              &       15.7            &       17.0    &22.5/20.5/16.0$^c$\\
                $       i^\prime $                              &       15.7            &       17.0    &22.5/20.5/16.0 \\
                $       J$                              &       17.7            &       19.0    &20.5/19.7/18.4 \\
                $       H$                              &       16.0            &       17.3    &19.5/18.7/17.4 \\
                $       K_s$                              &       15.7            &       17.0    &18.3/17.7/16.4 \\

        \hline
        \end{tabular}\\
        $^a$: for 0.5 arcsec slit.
        $^b$: for 90\% telescope reflectivity, 1 hr exposure, 1.0 arcsec slit, 1 arcsec seeing and SNR=10.\\
        $^c$: 1hr/10min/1 min exposure.
        \end{table*}

	\section{Conclusion} \label{ConclusionSection}
	
 The TANSPEC is a unique instrument that has the ability to deliver simultaneous optical-NIR (0.55-2.5 $\mu$m) spectrum in a single exposure. The TANSPEC was delivered and commissioned on the 3.6-m DOT successfully. It has been used for both commissioning tests and science observations during 2019-2020. The two spectroscopic modes of the TANSPEC, XD mode and prism mode, provide spectra with resolutions R$\sim$2700 and R$\sim$100-350, respectively. The main parameters of the TANSPEC obtained during commissioning tests and science observations are tabulated in Table \ref{sens}. It was officially released to the astronomy community in October 2020 and is currently being used heavily for science observations by many astronomers worldwide.

	\subsection* {Acknowledgments}
	We thank the staff of the  3.6-m DOT and 1.3-m DFOT, Devasthal (ARIES), for their co-operation during the observations. It is a pleasure to thank the
	members of the IR astronomy group (Department of Astronomy and Astrophysics) at TIFR for their support during observations. 	
	We acknowledge the \doclink{http://irtfweb.ifa.hawaii.edu/~spex/overview/overview.html}{Spex instrument} which we relied heavily on for the initial concept and the many practical lessons learned from that instrument.
	The Project In charge (PI) and Co-PI of TANSPEC project are thankful to the funding agencies 
(Department of Science and Technology (DST) and Department of Atomic Energy (DAE), Government of India, India) for approving the project and Directors of TIFR 
and ARIES for their support and encouragement. Thanks to the ARIES Governing council and Project Management Board (PMB) (3.6-m DOT) 
for reviewing time-to-time the progress of the project, and finally, the DOT and TIFR teams for their support 
during installation and commissioning of the TANSPEC on 3.6-m DOT. Finally, TIFR team acknowledges the support of the DAE, Government of India, under project Identification No. RTI 4002.

\section*{References}

\bibliographystyle{iopart-num}   
\bibliography{tanspec}   






\end{document}